\documentclass[twocolumn,showpacs,preprintnumbers,amsmath,amssymb,superscriptaddress]{revtex4-1}
\usepackage{chemformula} % Formula subscripts using \ch{}
\usepackage{graphicx}
\usepackage[colorlinks=true,linkcolor=black,citecolor=blue]{hyperref}

\begin{document}

\title{Room temperature cavity polaritons with 3D hybrid perovskite - Towards low cost polaritonic devices}

\author{P. Bouteyre}
\affiliation{Laboratoire Aim\'e Cotton, CNRS, Univ. Paris-Sud, ENS Paris-Saclay, Universit\'e Paris-
Saclay, 91405 Orsay Cedex, France}
\author{H. S. Nguyen}
\affiliation{Universit\'e de Lyon, Insitut des Nanotechnologies de Lyon - INL, UMR CNRS 5270, CNRS, Ecole Centrale de Lyon, Ecully, F-69134, France}
\author{J.-S. Lauret}
\affiliation{Laboratoire Aim\'e Cotton, CNRS, Univ. Paris-Sud, ENS Paris-Saclay, Universit\'e Paris-
Saclay, 91405 Orsay Cedex, France}
\author{G. Allard-Tripp\'e}
\affiliation{Laboratoire Aim\'e Cotton, CNRS, Univ. Paris-Sud, ENS Paris-Saclay, Universit\'e Paris-
Saclay, 91405 Orsay Cedex, France}
\author{G. Delport}
\affiliation{Laboratoire Aim\'e Cotton, CNRS, Univ. Paris-Sud, ENS Paris-Saclay, Universit\'e Paris-
Saclay, 91405 Orsay Cedex, France}
\author{F. L\'ed\'ee}
\affiliation{Laboratoire Aim\'e Cotton, CNRS, Univ. Paris-Sud, ENS Paris-Saclay, Universit\'e Paris-
Saclay, 91405 Orsay Cedex, France}
\author{H. Diab}
\affiliation{Laboratoire Aim\'e Cotton, CNRS, Univ. Paris-Sud, ENS Paris-Saclay, Universit\'e Paris-
Saclay, 91405 Orsay Cedex, France}
\author{A. Belarouci}
\affiliation{Universit\'e de Lyon, Insitut des Nanotechnologies de Lyon - INL, UMR CNRS 5270, CNRS, Ecole Centrale de Lyon, Ecully, F-69134, France}
\author{C. Seassal}
\affiliation{Universit\'e de Lyon, Insitut des Nanotechnologies de Lyon - INL, UMR CNRS 5270, CNRS, Ecole Centrale de Lyon, Ecully, F-69134, France}
\author{D. Garrot}
\affiliation{Groupe d$^\prime$Etude de la Mati\`ere Condens\'ee, Universit\'e de Versailles Saint Quentin En
Yvelines, Universit\'e Paris-Saclay, 45 Avenue des Etats-Unis, 78035, Versailles, France}
\author{F. Bretenaker}
\affiliation{Laboratoire Aim\'e Cotton, CNRS, Univ. Paris-Sud, ENS Paris-Saclay, Universit\'e Paris-
Saclay, 91405 Orsay Cedex, France}
\author{E. Deleporte }
\email{emmanuelle.deleporte@ens-cachan.fr}
\affiliation{Laboratoire Aim\'e Cotton, CNRS, Univ. Paris-Sud, ENS Paris-Saclay, Universit\'e Paris-
Saclay, 91405 Orsay Cedex, France}

\date{\today}

\begin{abstract}
Hybrid halide perovskites are now considered as key materials for contemporary research in photovoltaics and nanophotonics. In particular, because these materials can be solution processed, they represent a great hope for obtaining low cost devices. While the potential of 2D layered hybrid perovskites for polaritonic devices operating at room temperature has been demonstrated in the past, the potential of the 3D perovskites has been much less explored for this particular application. Here, we report the strong exciton-photon coupling with 3D bromide hybrid perovskite. Cavity polaritons are experimentallly demonstrated from both reflectivity and photoluminescence experiments, at room temperature, in a 3$\lambda$/2 planar microcavity containing a large surface spin-coated \ch{CH_3NH_3PbBr_3} thin film. A microcavity quality factor  of 92 was found and a large Rabi splitting of 70 meV was measured. This result paves the way to low-cost polaritonic devices operating at room temperature, potentially electrically injectable as 3D hybrid perovskites present good transport properties. 
\end{abstract}


\maketitle

Cavity polaritons are half-light half-matter quasi-particles arising from the strong coupling regime between excitonic and photonic modes \cite{Weisbuch1992}. Such regime is achieved when the coupling strength, related to the oscillator strength quantifying the light-matter interaction in a material, is larger than the dissipation rates of uncoupled excitons and cavity photons. Thanks to its hybrid nature, cavity polaritons inherit the best features of both the excitonic and photonic component: strongly nonlinear bosonic particles which can propagate balistically over macroscopic distance, and can be injected/probed via optical means. These fascinating properties suggest not only a playground for studying physics of out of equilibrium Bose Einstein condensation, but also a potential platform for all-optical devices. In the later direction, many proof-of-concepts of polaritonic devices have been reported:  polaritonic lasers \cite{Kasprzak2006}, polariton transistors \cite{gao2012}, resonant tunnelling diodes \cite{Nguyen2013}, interferometer \cite{Sturm2014}, optical gates \cite{gao2012}, and optical router \cite{Marsault2015}. Most of these demonstrations are in GaAs-based system – the most accomplished technologies to engineer cavity polaritons. However, due to the small excitonic effects and oscillator strength in GaAs, their operating regime is limited to cryogenic temperature. For this reason, materials presenting strong excitonic effects at room temperature, such as the high band gap materials GaN \cite{Semond2005,Christopoulos2007} or ZnO \cite{Sturm2009,Li2013a} are actively studied. However, the achievement of inorganic semiconductor engineered confined microstructures need sophisticated and high temperature epitaxial techniques. Looking for low-cost solutions, soft chemistry and low temperature processed materials presenting strong excitonic effects were also considered. The strong coupling regime at room temperature has been demonstrated in planar microcavities containing organic materials \cite{lidzey1998strong,kena-cohen2010,Daskalakis2014,Plumhof2013} or organic-inorganic halide perovskites such as \ch{(C_6H_5C_2H_4NH_3)_2PbI_4} \cite{brehier2006strong,lanty2008strong,han2012high,Fujita1998,nguyen2014quantum,Wang2018g}, taking advantage of very stable excitons and strong oscillator strengths. Nevertheless, the poor transport properties of these materials reveals as a major drawback for the electrical injection of polariton-based emitting devices. Most recently, monolayers of transitional metal dichalcogenides has joined the list of materials for room temperature polaritons thanks to prominent excitonic features \cite{Dufferwiel2015,Liu2015}. However, the fragility of these 2D materials prevents their use in large scale applications. As consequence, the quest for an ideal candidate for room temperature polaritons is still opened: a homogeneous large surface thin film material, for such devices are multi-layered structures, with good transport mobilities for electrical injection, stable Wannier excitons at room temperature,  large exciton oscillator strength and a low-cost method for synthesis, deposition and patterning into microstructures.\\ 

\begin{figure*}
  \includegraphics[width=0.8\textwidth]{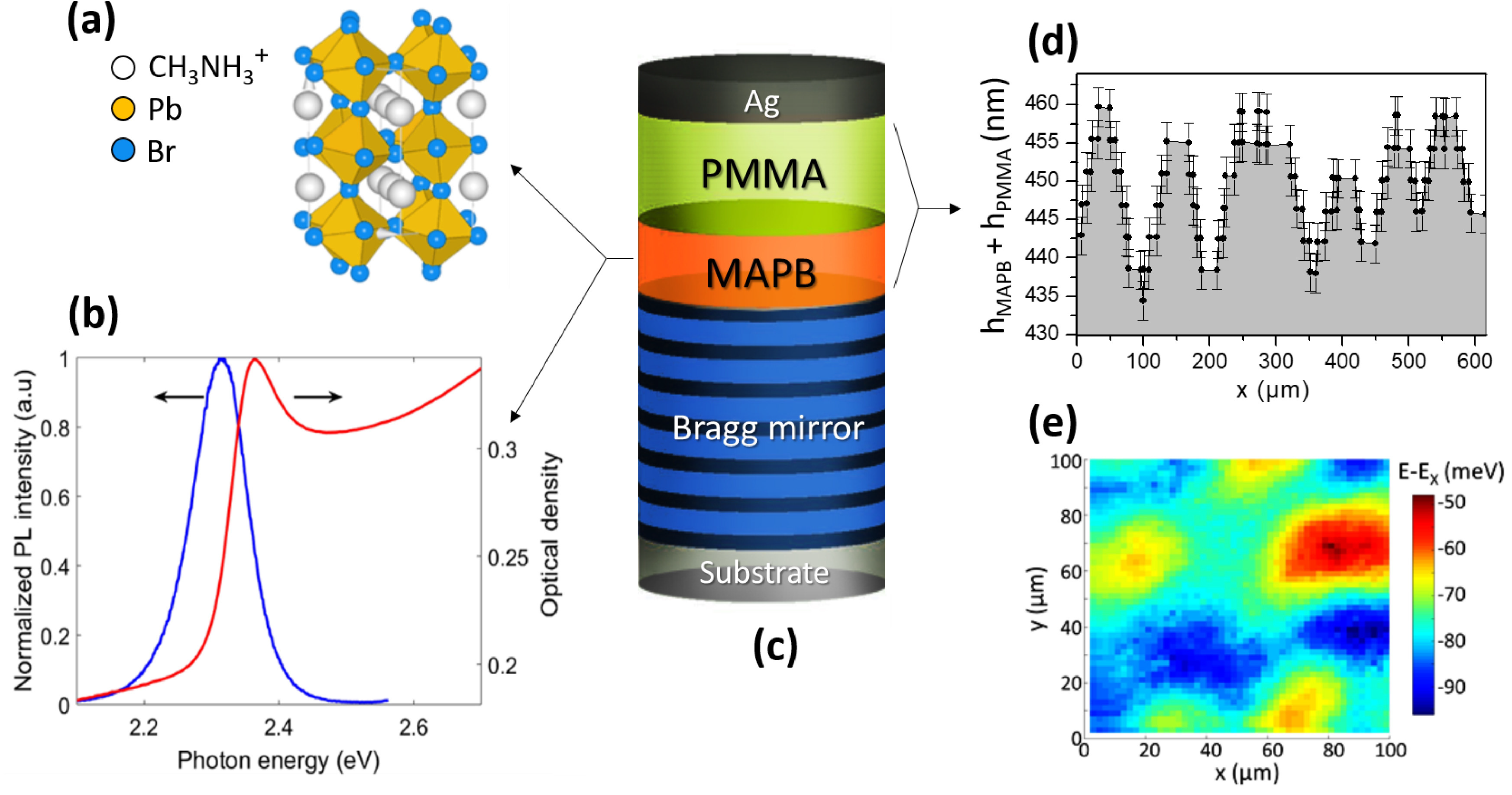}
  \caption{MAPB-based microcavity structure and its characterization. a) MAPB crystal structure b) Absorption and Photoluminescence (PL) on the MAPB thin film. c) Sketch of the microcavity.  d) Profilometry scan of the MAPB/PMMA layer before the silver deposition d) Spatial map of the microcavity detuning variation due to the MAPB/PMMA roughness measured with a $\mu$-PL setup.  }
  \label{Figure1}
\end{figure*}

Among the organic-inorganic halide perovskites, the ones having the basic chemical formula \ch{CH_3NH_3PbX_3}, named 3D perovskites, where X is a halogen atom (I, Br or Cl or a mixing of these atoms) has emerged recently, first in the framework of photovoltaics \cite{NREL} and then in the framework of emitting devices such as electroluminescent diodes \cite{Kim2014,Tan2014,Xing2014} and lasers \cite{deschler2014high,zhu2015lead,Chen2018c,Zhang2016b}. In these materials, the carriers can move along the 3 directions of the space, with a relatively good mobility of the order of 10 $cm^2.V^{-1}.S^{-1}$ \cite{Ponseca2014,Xing2013} which could be favourable for the electrical injection of polaritonic devices. The stability of the excitonic properties at room temperature depends on the nature of the halogen atom. In the iodide perovskite \ch{CH_3NH_3PbI_3}, the consensus is now that the free carriers are predominantly formed by photo-excitation at room temperature, which explains the high Power Conversion Efficiencies of the solar cells \cite{miyata2015direct,Phuong2016,Yang2017a}, but prevents to observe the strong coupling regime. Changing the atom I by Br or Cl, more stable excitons can be obtained. Large exciton binding energy values from 41 to 75 meV, larger than $k_BT$ at room temperature, are found for chloride based perovskites \cite{Zhang2016b,Comin2015,Saba2015,Yamada2018,Protesescu2015} and in fact the strong coupling regime at room temperature could be seen very recently in \ch{CsPbCl_3} large nanoplatelets of micrometric size \cite{su2017room}. Various values from 15 to 110 meV can be found in the literature for bromide based perovskites \cite{Chen2018c,Comin2015,Saba2015,tanaka2003comparative,Yang2015,soufiani2015polaronic,Galkowski2016,tilchin2016hydrogen,Wu2016,wolf2017structural,Niesner2017,Droseros2018}, and to the best of our knowledge, strong coupling has been obtained only in confined structures such as \ch{CsPbBr_3} nanowires \cite{Park2016,evans2018continuous} or \ch{CH_3NH_3PbBr_3} nanowires \cite{Zhang2017}, in which the exciton binding energy is strengthened by confinement effects.\\ 
In this letter, we report for the first time the strong exciton-photon coupling regime, at room temperature, in a planar microcavity containing a large surface of spin-coated \ch{CH_3NH_3PbBr_3} thin film as active material. Such experimental demonstration has multifold signification. From a material point of view, the observation of cavity polaritons is coherent with a dominant excitonic regime in bromide 3D perovskite at room temperature. Moreover, from application point of view, this result opens the door towards low cost room temperature polaritonic devices which can be electrically injected.\\

\begin{figure*}
  \includegraphics[width=0.8\textwidth]{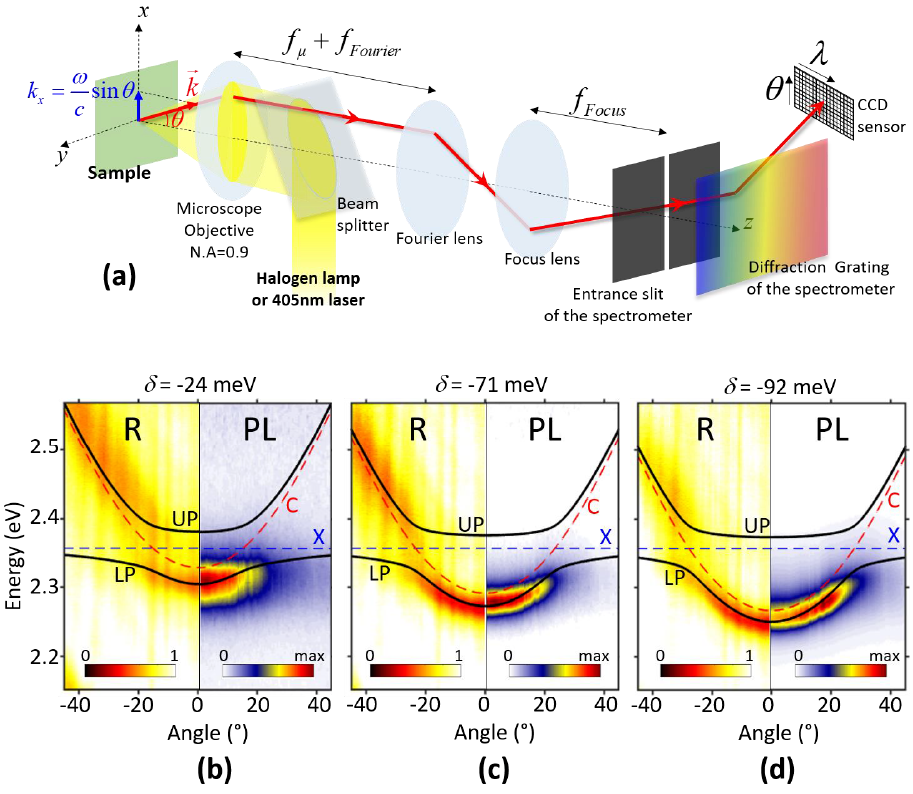}
  \caption{Angle resolved photoluminescence and reflectivity (ARPL and ARR) by Fourier spectroscopy. a) Sketch of the Fourier spectroscopy set-up. b) c) and d) ARPL and ARR of the microcavity at three different detuning: b) $\delta$= -24 meV ,  c)- 71 meV , d) - 92 meV. }
  \label{Figure2}
\end{figure*}

Our sample consists of a perovskite \ch{CH_3NH_3PbBr_3} layer, called hereafter MAPB, which is embedded in a 3$\lambda$/2 P\'erot-Fabry microcavity (see Figure \ref{Figure1}c ). The bottom mirror is a half an inch commercial dielectric mirror (LayerTec Corp.), centred at 2.4 eV at $8^{\circ}$ of incidence, whose reflectivity at 2.4 eV is 97.74\% and whose stop band extends from 2.1 eV to 2.78 eV. A 100 nm thick layer of MAPB perovskite, whose crystal structure is shown in Figure \ref{Figure1}a, is deposited directly on top of this dielectric mirror. We have optimized the crystallinity of the thin film, using a 2 step method inspired by Cadelano et al. \cite{cadelano2015can}: a thin film of \ch{PbBr_2} is spin-coated on top of the dielectric mirror and is later immersed in a \ch{CH_3NH_3Br} solution, the reaction between the two species forms the perovskite thin film. The absorption and photoluminescence (PL) spectra of this thin layer of MAPB is shown in Figure \ref{Figure1}b: the exciton energy lies at 2.35 eV in the absorption spectrum. Then a 350 nm thick PMMA (Poly(methyl)methacrylate) thin film is deposited by spin-coating as a spacer layer to tune the cavity photon mode close to the exciton energy. Finally, the top mirror of the microcavity is produced by thermal evaporation of a 30 nm silver layer directly on the PMMA layer. More details on the microcavity fabrication method is given in Supplementary section 1.\\

Surface profilometry measurements performed on the MAPB/PMMA layer before the silver deposition (Figure \ref{Figure1}d) show a total average thickness of 450 nm with around 30 nm of roughness and grains of 25 to 50 $\mu m$ of width. To confirm the cavity length variation measured by profilometry, spatial resolved photoluminescence (PL) of the microcavity is performed using a $\mu$-PL set-up with a femtosecond pulsed laser (100fs, 80 MHz, 405 nm) excitation. The cavity detuning is defined by $\delta=E_0-E_X$, with $E_0$ the cavity mode energy at normal incidence and $E_X=2.35$ eV the exciton energy (from the absorption spectrum in Figure \ref{Figure1}b). Figure \ref{Figure1}e presents the mapping of $\delta$ over 100x100$\mu m^2$, showing domains of constant detuning of size of 20-50 $\mu m$ grain. The agreement between profilometry and PL measurement provides a very good estimation of the inhomogeneous broadening of our microcavity. \\

Angle resolved reflectivity and photoluminescence (ARR and ARPL) are performed with a Fourier spectroscopy set-up imaging the photoluminescence in the momentum-space (Figure \ref{Figure2}a). The microcavity is excited by a white light (Halogen lamp) with a spot-size of 10 $\mu m^2$ for ARR experiment, and by a picosecond pulsed laser (50ps, 80 MHz, 405nm), with a spot-size of 1 $\mu m^2$ for ARPL measurements. We note that in both experiment, the spot-size is small enough to probe the cavity characteristics with no inhomogeneous effect caused by the roughness. Figure \ref{Figure2} b, c and d show the angle-resolved reflectivity (ARR, left panel) and angle-resolved photoluminescence (ARPL, right panel) maps for three different detunings of the microcavity: $\delta$ = -24 meV, - 71 meV, - 92 meV when probing on different positions of the sample.\\

\begin{figure*}
  \includegraphics[width=0.8\textwidth]{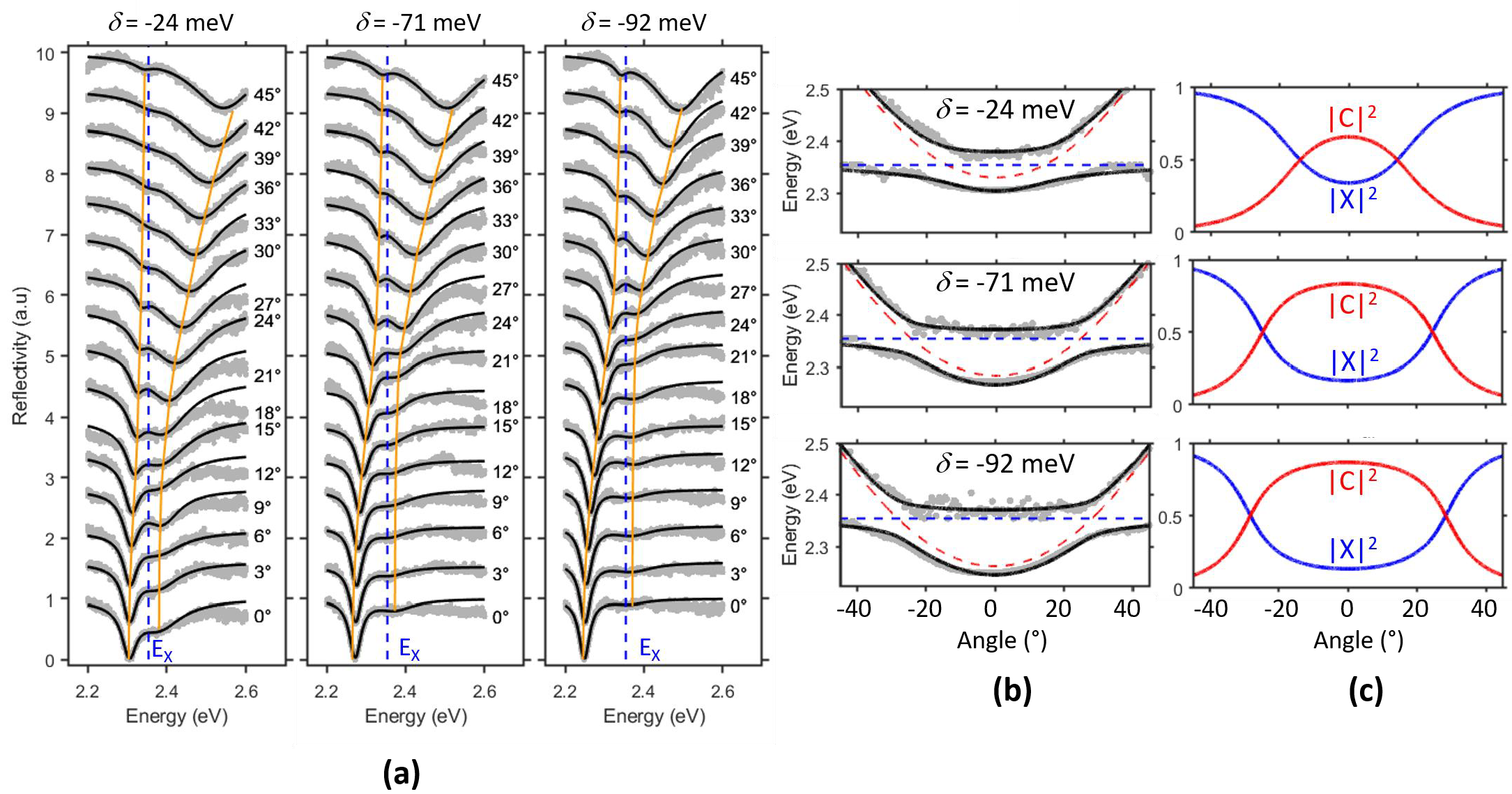}
  \caption{Analysis of the ARR results a) Slices of the reflectivity maps at different angles fitted with two lorentzian functions. Yellow lines are guides for the eyes. b) Reflectivity dispersions fitted with the two level model (solid lines). Dotted lines correspond to the dispersion of the bare photon (red) and excitonic (blue) modes c) Photonic (red line) and excitonic (blue line) weights of the lower energy polariton as a function of the angle at the three different cavity detunings.}
  \label{Figure3}
\end{figure*}

We will first discuss the experimental results obtained from ARR measurements. Two dispersion branches can be seen in the reflectivity parts of each image. The energy position, intensity and linewidth of the two branches are angle dependent. Most importantly, a clear anti-crossing between these two branches is observed, which is the signature of a strong-coupling regime between the exciton mode and the photon mode. These two dispersion branches, called upper polariton branch (UPB) and lower polariton branch (LPB) can be calculated from the bare cavity and bare exciton dispersion by using a standard two-level model of Hamiltonian given by:

\begin{equation}
\label{eq1}
H= 
\begin{pmatrix}
E_{ph}(\theta) -i\gamma_{ph} & V \\
V  & E_{X} -i\gamma_{X}
\end{pmatrix}
\end{equation}

where $E_{ph}(\theta)$ and $E_X$ are energies of the photonic mode and excitonic mode; V is the exciton-photon coupling strength; $\gamma_{ph}$ is the cavity linewidth, related to its quality factor, and $\gamma_X$ is the exciton linewidth. Apart from the energy of the photonic mode, all other parameters are considered independent of $\theta$ (the excitonic dispersion is considered flat since the excitonic mass is much larger than the photonic mass). The dispersion of the photonic mode $E_{ph}$ is given by \cite{Lidzey2002}:

\begin{equation}
\label{eq2}
\centering
E_{ph}(\theta)=\frac{E_X+\delta}{\sqrt{1-\frac{sin^2(\theta)}{n_{eff}^2}}}
\end{equation}

where $n_{eff}$ is the effective refractive index of the entire cavity.\\
 
     The resolution of the two-level model results to two eigenstates, coherent superposition of the excitonic and photonic modes with the following eigenvalues:
 
\begin{align}
\label{eq3}
\centering
E_{UPB,LPB}&(\theta)=\frac{1}{2}[E_{ph}(\theta)+E_{X}-i(\gamma_{ph}+\gamma_{X})]\\ \nonumber
&\pm\sqrt{V^2+\frac{1}{4}[E_{X}-E_{ph}(\theta)+i(\gamma_{ph}-\gamma_{X})]^2}
\end{align}

where the real parts of  $E_{(UPB,LPB)}$ correspond respectively to the UPB and LPB energies, and the imaginary parts correspond to their linewidths. In order to estimate precisely the photonic and excitonic parameters, reflectivity spectrum of each angle (i.e. slices of the reflectivity maps) is fitted with a double lorentzian function (see Figure \ref{Figure3}a). The peak positions of each angle are reported in Figure \ref{Figure3}b in a dispersion diagram and fitted with the real part of eq \ref{eq3} where $\delta$, $n_{eff}$, $E_X$, V, $\gamma_{ph}$ and $\gamma_X$ are let as free parameters. As shown in Figure \ref{Figure3}b, the experimental data of three different detunings is reproduced perfectly by the two-level model with the same set of parameters: $n_{eff}$ =1.7, $E_X$ =2.355 eV, V= 50 meV, $\gamma_{ph}$ =25 meV and $\gamma_X$= 90 meV. The value obtained for $E_X$ is in agreement with the energy position of the perovskite absorption peak (Figure \ref{Figure3}b) and the one reported in the literature \cite{soufiani2015polaronic, tilchin2016hydrogen, tanaka2003comparative, Yang2015, Wu2016}. The value of $n_{eff}$ is very close to the cavity effective refractive index of 1.742 (see Supplementary Section 3). The Rabi splitting is found to be 70 meV, which is higher than the mean value of the cavity and exciton linewidths $(\Omega_{Rabi}>1/2(\gamma_{ph}+\gamma_X))$, necessary condition to observe the strong coupling. Figure \ref{Figure3}c shows the calculation of the photonic weights $|C|^2$ and the excitonic weights $|X|^2$ of the LPB in function of the angle for the three different cavity detunings. We note that the hybrid nature of the UPB and LPB varies from a photonic-like behaviour to an exciton-like behaviour in function of the spectral distance between the cavity mode and the exciton energy: i) far from the anticrossing point, the hybridization is unbalanced and the LPB (UPB) is either mostly photonic(excitonic) or excitonic(photonic) ii) the hybridization becomes more balanced when getting closer to the anticrossing point where photons and excitons are mixed equally.\\ 

\begin{figure}
  \includegraphics[width=0.4\textwidth]{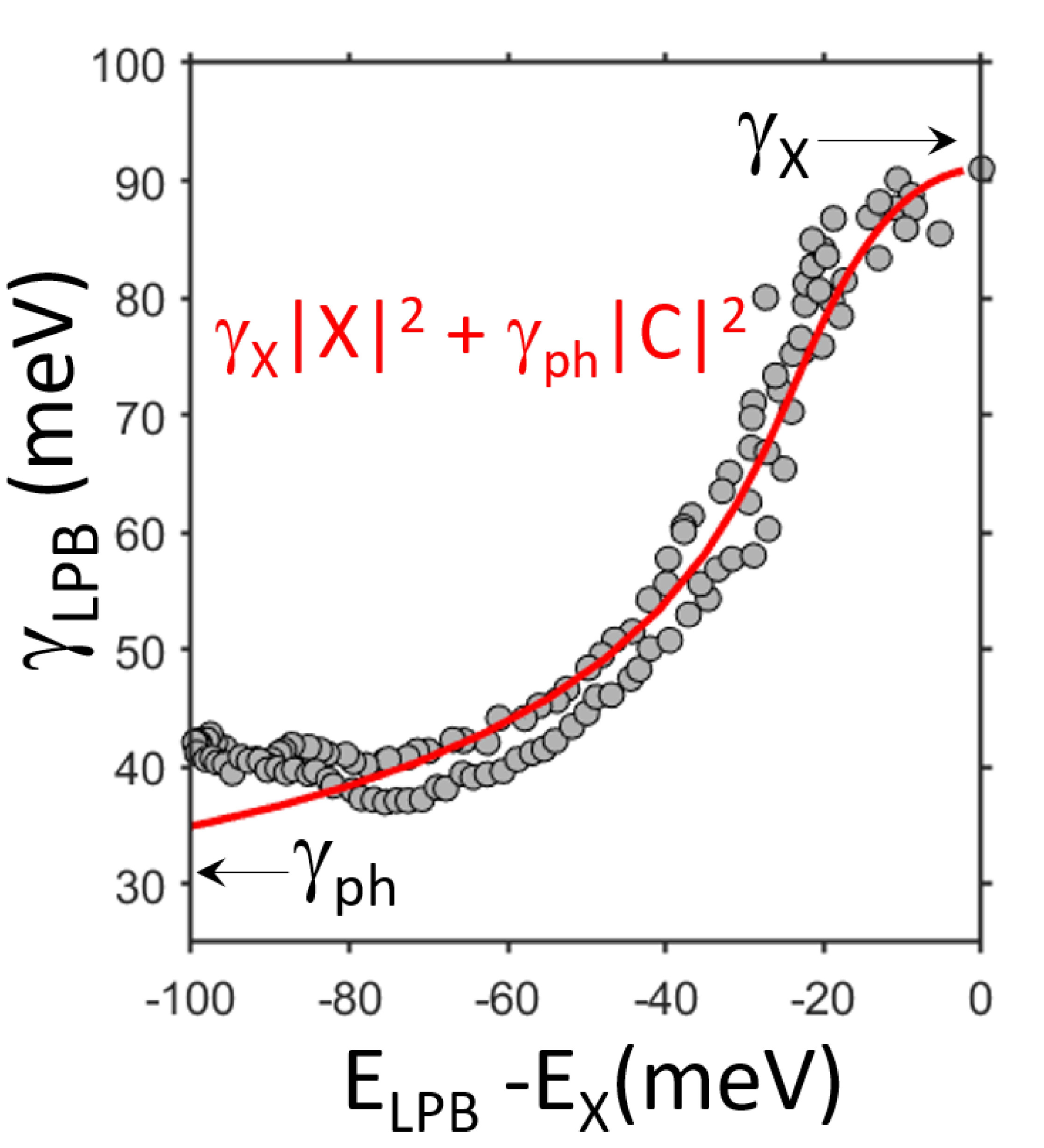}
  \caption{Analysis of the ARPL results. Photoluminescence linewidth in function of the emission angle, plotted here as a function of the difference between the exciton and emission energies $(E_{LP}-E_X)$ in the case of the cavity detuning $\delta$ = -92 meV.}
  \label{Figure4}
\end{figure}

In the following, we will discuss the ARPL results reported in Figure \ref{Figure2}. Emission from the LPB is clearly observed and in good agreement with ARR measurements and the analytical fitting. Typical to polaritonic system at room temperature with high bandgap materials \cite{Flatten2016,Virgili2011,Christmann2008a,Lai2013}, the PL of the UPB is not observed. Note that for the large detuning $\delta$ = -92 meV, the emission accumulates next to the inflexion point of the LPB at around $20^{\circ}$, possibly due to the bottleneck effect \cite{Lai2013,Tassone1997}. As it is often the case in systems presenting a large Rabi splitting (in GaN and ZnO for example \cite{Zuniga-Perez2014}) the curve inflexion of the LPB, characteristic of the polaritonic emission, cannot be observed in the experimental photoluminescence.\\

In order to bring further evidence of the strong coupling regime from ARPL, the study of the PL dispersion linewidth has been made Indeed, from eq.\ref{eq3}, the linewidth of the LPB is given by: 

\begin{equation}
\label{eq4}
\centering
\gamma_{LPB}=|C|^2 \gamma_{ph}+|X|^2 \gamma_X
\end{equation}

thus the linewidth of LPB will provide an important information about the hybrid nature of this state. Figure \ref{Figure4} shows the PL linewidth of the LPB, in the case of the cavity detuning $\delta$ = -92 meV, in function of the difference between the exciton and emission energies $(E_{LPB}-E_X)$, which varies with the emission angle as the emission energy varies with the angle. The experimental data is perfectly fitted with the same parameters of the values obtained from the ARR:  $\gamma_X$=91 meV and $\gamma_{ph}$=25 meV, giving the cavity quality factor to be 92. Most importantly, while scanning $E_{LPB}-E_X$, the polariton linewidth varies continuously from the excitonic to the photonic linewidth, strongly confirms the hybrid nature half-matter/half-light of cavity polaritons.\\
 
 In conclusion, we have demonstrated in both reflectivity and photoluminescence experiments a strong coupling regime in a \ch{CH_3NH_3PbBr_3} large surface spin-coated thin film-based 3$\lambda$/2 microcavity at room temperature. The microcavity quality factor was found to be around 92 and the measured Rabi splitting $\Omega_{Rabi}$ was 70 meV. Firstly, our observation of the strong coupling regime is coherent with a dominant excitonic regime in \ch{CH_3NH_3PbBr_3} at room temperature, in good agreement with the most recent paper about this topic \cite{Droseros2018}. Secondly, this result is particularly important because it indicates that it is possible to obtain the strong coupling regime in solution-processed films of 3D perovskite. The versatility of the solution processed fabrication techniques, make them very attractive for integrable devices, because they are low cost and low temperature technologies, allowing to have large surfaces. This strong coupling regime gives then hope towards 3D hybrid perovskite-based low cost polaritonic devices, such as low threshold polariton lasers operating at room-temperature by electrical injection due to the good electronic mobility properties of these perovskites.\\

\noindent Acknowledgement : \\

This work is supported by Agence Nationale de la Recherche (ANR) within the projects POPEYE and EMIPERO. The work of P.Bouteyre is supported by the Direction G\'en\'erale de l$^\prime$Armement (DGA).\\
We thank Rasta Ghasemi, from Institut d$^\prime$Alembert de l$^\prime$Ecole Normale Sup\'erieure de Cachan, for her technical support in the evaporation of the metallic mirrors. \\

\newpage

\bibliography{mybib}

%merlin.mbs apsrev4-1.bst 2010-07-25 4.21a (PWD, AO, DPC) hacked
%Control: key (0)
%Control: author (8) initials jnrlst
%Control: editor formatted (1) identically to author
%Control: production of article title (-1) disabled
%Control: page (0) single
%Control: year (1) truncated
%Control: production of eprint (0) enabled
\begin{thebibliography}{60}%
\makeatletter
\providecommand \@ifxundefined [1]{%
 \@ifx{#1\undefined}
}%
\providecommand \@ifnum [1]{%
 \ifnum #1\expandafter \@firstoftwo
 \else \expandafter \@secondoftwo
 \fi
}%
\providecommand \@ifx [1]{%
 \ifx #1\expandafter \@firstoftwo
 \else \expandafter \@secondoftwo
 \fi
}%
\providecommand \natexlab [1]{#1}%
\providecommand \enquote  [1]{``#1''}%
\providecommand \bibnamefont  [1]{#1}%
\providecommand \bibfnamefont [1]{#1}%
\providecommand \citenamefont [1]{#1}%
\providecommand \href@noop [0]{\@secondoftwo}%
\providecommand \href [0]{\begingroup \@sanitize@url \@href}%
\providecommand \@href[1]{\@@startlink{#1}\@@href}%
\providecommand \@@href[1]{\endgroup#1\@@endlink}%
\providecommand \@sanitize@url [0]{\catcode `\\12\catcode `\$12\catcode
  `\&12\catcode `\#12\catcode `\^12\catcode `\_12\catcode `\%12\relax}%
\providecommand \@@startlink[1]{}%
\providecommand \@@endlink[0]{}%
\providecommand \url  [0]{\begingroup\@sanitize@url \@url }%
\providecommand \@url [1]{\endgroup\@href {#1}{\urlprefix }}%
\providecommand \urlprefix  [0]{URL }%
\providecommand \Eprint [0]{\href }%
\providecommand \doibase [0]{http://dx.doi.org/}%
\providecommand \selectlanguage [0]{\@gobble}%
\providecommand \bibinfo  [0]{\@secondoftwo}%
\providecommand \bibfield  [0]{\@secondoftwo}%
\providecommand \translation [1]{[#1]}%
\providecommand \BibitemOpen [0]{}%
\providecommand \bibitemStop [0]{}%
\providecommand \bibitemNoStop [0]{.\EOS\space}%
\providecommand \EOS [0]{\spacefactor3000\relax}%
\providecommand \BibitemShut  [1]{\csname bibitem#1\endcsname}%
\let\auto@bib@innerbib\@empty
%</preamble>
\bibitem [{\citenamefont {Weisbuch}\ \emph {et~al.}(1992)\citenamefont
  {Weisbuch}, \citenamefont {Nishioka}, \citenamefont {Ishikawa},\ and\
  \citenamefont {Arakawa}}]{Weisbuch1992}%
  \BibitemOpen
  \bibfield  {author} {\bibinfo {author} {\bibfnamefont {C.}~\bibnamefont
  {Weisbuch}}, \bibinfo {author} {\bibfnamefont {M.}~\bibnamefont {Nishioka}},
  \bibinfo {author} {\bibfnamefont {A.}~\bibnamefont {Ishikawa}}, \ and\
  \bibinfo {author} {\bibfnamefont {Y.}~\bibnamefont {Arakawa}},\ }\href
  {\doibase 10.1103/physrevlett.69.3314} {\bibfield  {journal} {\bibinfo
  {journal} {Physical Review Letters}\ }\textbf {\bibinfo {volume} {69}},\
  \bibinfo {pages} {3314} (\bibinfo {year} {1992})}\BibitemShut {NoStop}%
\bibitem [{\citenamefont {Kasprzak}\ \emph {et~al.}(2006)\citenamefont
  {Kasprzak}, \citenamefont {Richard}, \citenamefont {Kundermann},
  \citenamefont {Baas}, \citenamefont {Jeambrun}, \citenamefont {Keeling},
  \citenamefont {Marchetti}, \citenamefont {Szyma{\'{n}}ska}, \citenamefont
  {Andr{\'{e}}}, \citenamefont {Staehli}, \citenamefont {Savona}, \citenamefont
  {Littlewood}, \citenamefont {Deveaud},\ and\ \citenamefont
  {Dang}}]{Kasprzak2006}%
  \BibitemOpen
  \bibfield  {author} {\bibinfo {author} {\bibfnamefont {J.}~\bibnamefont
  {Kasprzak}}, \bibinfo {author} {\bibfnamefont {M.}~\bibnamefont {Richard}},
  \bibinfo {author} {\bibfnamefont {S.}~\bibnamefont {Kundermann}}, \bibinfo
  {author} {\bibfnamefont {A.}~\bibnamefont {Baas}}, \bibinfo {author}
  {\bibfnamefont {P.}~\bibnamefont {Jeambrun}}, \bibinfo {author}
  {\bibfnamefont {J.~M.~J.}\ \bibnamefont {Keeling}}, \bibinfo {author}
  {\bibfnamefont {F.~M.}\ \bibnamefont {Marchetti}}, \bibinfo {author}
  {\bibfnamefont {M.~H.}\ \bibnamefont {Szyma{\'{n}}ska}}, \bibinfo {author}
  {\bibfnamefont {R.}~\bibnamefont {Andr{\'{e}}}}, \bibinfo {author}
  {\bibfnamefont {J.~L.}\ \bibnamefont {Staehli}}, \bibinfo {author}
  {\bibfnamefont {V.}~\bibnamefont {Savona}}, \bibinfo {author} {\bibfnamefont
  {P.~B.}\ \bibnamefont {Littlewood}}, \bibinfo {author} {\bibfnamefont
  {B.}~\bibnamefont {Deveaud}}, \ and\ \bibinfo {author} {\bibfnamefont
  {L.~S.}\ \bibnamefont {Dang}},\ }\href {\doibase 10.1038/nature05131}
  {\bibfield  {journal} {\bibinfo  {journal} {Nature}\ }\textbf {\bibinfo
  {volume} {443}},\ \bibinfo {pages} {409} (\bibinfo {year}
  {2006})}\BibitemShut {NoStop}%
\bibitem [{\citenamefont {Gao}\ \emph {et~al.}(2012)\citenamefont {Gao},
  \citenamefont {Eldridge}, \citenamefont {Liew}, \citenamefont {Tsintzos},
  \citenamefont {Stavrinidis}, \citenamefont {Deligeorgis}, \citenamefont
  {Hatzopoulos},\ and\ \citenamefont {Savvidis}}]{gao2012}%
  \BibitemOpen
  \bibfield  {author} {\bibinfo {author} {\bibfnamefont {T.}~\bibnamefont
  {Gao}}, \bibinfo {author} {\bibfnamefont {P.~S.}\ \bibnamefont {Eldridge}},
  \bibinfo {author} {\bibfnamefont {T.~C.~H.}\ \bibnamefont {Liew}}, \bibinfo
  {author} {\bibfnamefont {S.~I.}\ \bibnamefont {Tsintzos}}, \bibinfo {author}
  {\bibfnamefont {G.}~\bibnamefont {Stavrinidis}}, \bibinfo {author}
  {\bibfnamefont {G.}~\bibnamefont {Deligeorgis}}, \bibinfo {author}
  {\bibfnamefont {Z.}~\bibnamefont {Hatzopoulos}}, \ and\ \bibinfo {author}
  {\bibfnamefont {P.~G.}\ \bibnamefont {Savvidis}},\ }\href {\doibase
  10.1103/physrevb.85.235102} {\bibfield  {journal} {\bibinfo  {journal}
  {Physical Review B}\ }\textbf {\bibinfo {volume} {85}} (\bibinfo {year}
  {2012}),\ 10.1103/physrevb.85.235102}\BibitemShut {NoStop}%
\bibitem [{\citenamefont {Nguyen}\ \emph {et~al.}(2013)\citenamefont {Nguyen},
  \citenamefont {Vishnevsky}, \citenamefont {Sturm}, \citenamefont {Tanese},
  \citenamefont {Solnyshkov}, \citenamefont {Galopin}, \citenamefont
  {Lema{\^{\i}}tre}, \citenamefont {Sagnes}, \citenamefont {Amo}, \citenamefont
  {Malpuech},\ and\ \citenamefont {Bloch}}]{Nguyen2013}%
  \BibitemOpen
  \bibfield  {author} {\bibinfo {author} {\bibfnamefont {H.~S.}\ \bibnamefont
  {Nguyen}}, \bibinfo {author} {\bibfnamefont {D.}~\bibnamefont {Vishnevsky}},
  \bibinfo {author} {\bibfnamefont {C.}~\bibnamefont {Sturm}}, \bibinfo
  {author} {\bibfnamefont {D.}~\bibnamefont {Tanese}}, \bibinfo {author}
  {\bibfnamefont {D.}~\bibnamefont {Solnyshkov}}, \bibinfo {author}
  {\bibfnamefont {E.}~\bibnamefont {Galopin}}, \bibinfo {author} {\bibfnamefont
  {A.}~\bibnamefont {Lema{\^{\i}}tre}}, \bibinfo {author} {\bibfnamefont
  {I.}~\bibnamefont {Sagnes}}, \bibinfo {author} {\bibfnamefont
  {A.}~\bibnamefont {Amo}}, \bibinfo {author} {\bibfnamefont {G.}~\bibnamefont
  {Malpuech}}, \ and\ \bibinfo {author} {\bibfnamefont {J.}~\bibnamefont
  {Bloch}},\ }\href {\doibase 10.1103/physrevlett.110.236601} {\bibfield
  {journal} {\bibinfo  {journal} {Physical Review Letters}\ }\textbf {\bibinfo
  {volume} {110}} (\bibinfo {year} {2013}),\
  10.1103/physrevlett.110.236601}\BibitemShut {NoStop}%
\bibitem [{\citenamefont {Sturm}\ \emph {et~al.}(2014)\citenamefont {Sturm},
  \citenamefont {Tanese}, \citenamefont {Nguyen}, \citenamefont {Flayac},
  \citenamefont {Galopin}, \citenamefont {Lema{\^{\i}}tre}, \citenamefont
  {Sagnes}, \citenamefont {Solnyshkov}, \citenamefont {Amo}, \citenamefont
  {Malpuech},\ and\ \citenamefont {Bloch}}]{Sturm2014}%
  \BibitemOpen
  \bibfield  {author} {\bibinfo {author} {\bibfnamefont {C.}~\bibnamefont
  {Sturm}}, \bibinfo {author} {\bibfnamefont {D.}~\bibnamefont {Tanese}},
  \bibinfo {author} {\bibfnamefont {H.}~\bibnamefont {Nguyen}}, \bibinfo
  {author} {\bibfnamefont {H.}~\bibnamefont {Flayac}}, \bibinfo {author}
  {\bibfnamefont {E.}~\bibnamefont {Galopin}}, \bibinfo {author} {\bibfnamefont
  {A.}~\bibnamefont {Lema{\^{\i}}tre}}, \bibinfo {author} {\bibfnamefont
  {I.}~\bibnamefont {Sagnes}}, \bibinfo {author} {\bibfnamefont
  {D.}~\bibnamefont {Solnyshkov}}, \bibinfo {author} {\bibfnamefont
  {A.}~\bibnamefont {Amo}}, \bibinfo {author} {\bibfnamefont {G.}~\bibnamefont
  {Malpuech}}, \ and\ \bibinfo {author} {\bibfnamefont {J.}~\bibnamefont
  {Bloch}},\ }\href {\doibase 10.1038/ncomms4278} {\bibfield  {journal}
  {\bibinfo  {journal} {Nature Communications}\ }\textbf {\bibinfo {volume}
  {5}} (\bibinfo {year} {2014}),\ 10.1038/ncomms4278}\BibitemShut {NoStop}%
\bibitem [{\citenamefont {Marsault}\ \emph {et~al.}(2015)\citenamefont
  {Marsault}, \citenamefont {Nguyen}, \citenamefont {Tanese}, \citenamefont
  {Lema{\^{\i}}tre}, \citenamefont {Galopin}, \citenamefont {Sagnes},
  \citenamefont {Amo},\ and\ \citenamefont {Bloch}}]{Marsault2015}%
  \BibitemOpen
  \bibfield  {author} {\bibinfo {author} {\bibfnamefont {F.}~\bibnamefont
  {Marsault}}, \bibinfo {author} {\bibfnamefont {H.~S.}\ \bibnamefont
  {Nguyen}}, \bibinfo {author} {\bibfnamefont {D.}~\bibnamefont {Tanese}},
  \bibinfo {author} {\bibfnamefont {A.}~\bibnamefont {Lema{\^{\i}}tre}},
  \bibinfo {author} {\bibfnamefont {E.}~\bibnamefont {Galopin}}, \bibinfo
  {author} {\bibfnamefont {I.}~\bibnamefont {Sagnes}}, \bibinfo {author}
  {\bibfnamefont {A.}~\bibnamefont {Amo}}, \ and\ \bibinfo {author}
  {\bibfnamefont {J.}~\bibnamefont {Bloch}},\ }\href {\doibase
  10.1063/1.4936158} {\bibfield  {journal} {\bibinfo  {journal} {Applied
  Physics Letters}\ }\textbf {\bibinfo {volume} {107}},\ \bibinfo {pages}
  {201115} (\bibinfo {year} {2015})}\BibitemShut {NoStop}%
\bibitem [{\citenamefont {Semond}\ \emph {et~al.}(2005)\citenamefont {Semond},
  \citenamefont {Sellers}, \citenamefont {Natali}, \citenamefont {Byrne},
  \citenamefont {Leroux}, \citenamefont {Massies}, \citenamefont {Ollier},
  \citenamefont {Leymarie}, \citenamefont {Disseix},\ and\ \citenamefont
  {Vasson}}]{Semond2005}%
  \BibitemOpen
  \bibfield  {author} {\bibinfo {author} {\bibfnamefont {F.}~\bibnamefont
  {Semond}}, \bibinfo {author} {\bibfnamefont {I.~R.}\ \bibnamefont {Sellers}},
  \bibinfo {author} {\bibfnamefont {F.}~\bibnamefont {Natali}}, \bibinfo
  {author} {\bibfnamefont {D.}~\bibnamefont {Byrne}}, \bibinfo {author}
  {\bibfnamefont {M.}~\bibnamefont {Leroux}}, \bibinfo {author} {\bibfnamefont
  {J.}~\bibnamefont {Massies}}, \bibinfo {author} {\bibfnamefont
  {N.}~\bibnamefont {Ollier}}, \bibinfo {author} {\bibfnamefont
  {J.}~\bibnamefont {Leymarie}}, \bibinfo {author} {\bibfnamefont
  {P.}~\bibnamefont {Disseix}}, \ and\ \bibinfo {author} {\bibfnamefont
  {A.}~\bibnamefont {Vasson}},\ }\href {\doibase 10.1063/1.1994954} {\bibfield
  {journal} {\bibinfo  {journal} {Applied Physics Letters}\ }\textbf {\bibinfo
  {volume} {87}},\ \bibinfo {pages} {021102} (\bibinfo {year}
  {2005})}\BibitemShut {NoStop}%
\bibitem [{\citenamefont {Christopoulos}\ \emph {et~al.}(2007)\citenamefont
  {Christopoulos}, \citenamefont {von Högersthal}, \citenamefont {Grundy},
  \citenamefont {Lagoudakis}, \citenamefont {Kavokin}, \citenamefont
  {Baumberg}, \citenamefont {Christmann}, \citenamefont {Butt{\'{e}}},
  \citenamefont {Feltin}, \citenamefont {Carlin},\ and\ \citenamefont
  {Grandjean}}]{Christopoulos2007}%
  \BibitemOpen
  \bibfield  {author} {\bibinfo {author} {\bibfnamefont {S.}~\bibnamefont
  {Christopoulos}}, \bibinfo {author} {\bibfnamefont {G.~B.~H.}\ \bibnamefont
  {von Högersthal}}, \bibinfo {author} {\bibfnamefont {A.~J.~D.}\ \bibnamefont
  {Grundy}}, \bibinfo {author} {\bibfnamefont {P.~G.}\ \bibnamefont
  {Lagoudakis}}, \bibinfo {author} {\bibfnamefont {A.~V.}\ \bibnamefont
  {Kavokin}}, \bibinfo {author} {\bibfnamefont {J.~J.}\ \bibnamefont
  {Baumberg}}, \bibinfo {author} {\bibfnamefont {G.}~\bibnamefont
  {Christmann}}, \bibinfo {author} {\bibfnamefont {R.}~\bibnamefont
  {Butt{\'{e}}}}, \bibinfo {author} {\bibfnamefont {E.}~\bibnamefont {Feltin}},
  \bibinfo {author} {\bibfnamefont {J.-F.}\ \bibnamefont {Carlin}}, \ and\
  \bibinfo {author} {\bibfnamefont {N.}~\bibnamefont {Grandjean}},\ }\href
  {\doibase 10.1103/physrevlett.98.126405} {\bibfield  {journal} {\bibinfo
  {journal} {Physical Review Letters}\ }\textbf {\bibinfo {volume} {98}}
  (\bibinfo {year} {2007}),\ 10.1103/physrevlett.98.126405}\BibitemShut
  {NoStop}%
\bibitem [{\citenamefont {Sturm}\ \emph {et~al.}(2009)\citenamefont {Sturm},
  \citenamefont {Hilmer}, \citenamefont {Schmidt-Grund},\ and\ \citenamefont
  {Grundmann}}]{Sturm2009}%
  \BibitemOpen
  \bibfield  {author} {\bibinfo {author} {\bibfnamefont {C.}~\bibnamefont
  {Sturm}}, \bibinfo {author} {\bibfnamefont {H.}~\bibnamefont {Hilmer}},
  \bibinfo {author} {\bibfnamefont {R.}~\bibnamefont {Schmidt-Grund}}, \ and\
  \bibinfo {author} {\bibfnamefont {M.}~\bibnamefont {Grundmann}},\ }\href
  {\doibase 10.1088/1367-2630/11/7/073044} {\bibfield  {journal} {\bibinfo
  {journal} {New Journal of Physics}\ }\textbf {\bibinfo {volume} {11}},\
  \bibinfo {pages} {073044} (\bibinfo {year} {2009})}\BibitemShut {NoStop}%
\bibitem [{\citenamefont {Li}\ \emph {et~al.}(2013)\citenamefont {Li},
  \citenamefont {Orosz}, \citenamefont {Kamoun}, \citenamefont {Bouchoule},
  \citenamefont {Brimont}, \citenamefont {Disseix}, \citenamefont {Guillet},
  \citenamefont {Lafosse}, \citenamefont {Leroux}, \citenamefont {Leymarie},
  \citenamefont {Mexis}, \citenamefont {Mihailovic}, \citenamefont
  {Patriarche}, \citenamefont {R{\'{e}}veret}, \citenamefont {Solnyshkov},
  \citenamefont {Zuniga-Perez},\ and\ \citenamefont {Malpuech}}]{Li2013a}%
  \BibitemOpen
  \bibfield  {author} {\bibinfo {author} {\bibfnamefont {F.}~\bibnamefont
  {Li}}, \bibinfo {author} {\bibfnamefont {L.}~\bibnamefont {Orosz}}, \bibinfo
  {author} {\bibfnamefont {O.}~\bibnamefont {Kamoun}}, \bibinfo {author}
  {\bibfnamefont {S.}~\bibnamefont {Bouchoule}}, \bibinfo {author}
  {\bibfnamefont {C.}~\bibnamefont {Brimont}}, \bibinfo {author} {\bibfnamefont
  {P.}~\bibnamefont {Disseix}}, \bibinfo {author} {\bibfnamefont
  {T.}~\bibnamefont {Guillet}}, \bibinfo {author} {\bibfnamefont
  {X.}~\bibnamefont {Lafosse}}, \bibinfo {author} {\bibfnamefont
  {M.}~\bibnamefont {Leroux}}, \bibinfo {author} {\bibfnamefont
  {J.}~\bibnamefont {Leymarie}}, \bibinfo {author} {\bibfnamefont
  {M.}~\bibnamefont {Mexis}}, \bibinfo {author} {\bibfnamefont
  {M.}~\bibnamefont {Mihailovic}}, \bibinfo {author} {\bibfnamefont
  {G.}~\bibnamefont {Patriarche}}, \bibinfo {author} {\bibfnamefont
  {F.}~\bibnamefont {R{\'{e}}veret}}, \bibinfo {author} {\bibfnamefont
  {D.}~\bibnamefont {Solnyshkov}}, \bibinfo {author} {\bibfnamefont
  {J.}~\bibnamefont {Zuniga-Perez}}, \ and\ \bibinfo {author} {\bibfnamefont
  {G.}~\bibnamefont {Malpuech}},\ }\href {\doibase
  10.1103/physrevlett.110.196406} {\bibfield  {journal} {\bibinfo  {journal}
  {Physical Review Letters}\ }\textbf {\bibinfo {volume} {110}} (\bibinfo
  {year} {2013}),\ 10.1103/physrevlett.110.196406}\BibitemShut {NoStop}%
\bibitem [{\citenamefont {Lidzey}\ \emph {et~al.}(1998)\citenamefont {Lidzey},
  \citenamefont {Bradley}, \citenamefont {Skolnick}, \citenamefont {Virgili},
  \citenamefont {Walker},\ and\ \citenamefont {Whittaker}}]{lidzey1998strong}%
  \BibitemOpen
  \bibfield  {author} {\bibinfo {author} {\bibfnamefont {D.~G.}\ \bibnamefont
  {Lidzey}}, \bibinfo {author} {\bibfnamefont {D.}~\bibnamefont {Bradley}},
  \bibinfo {author} {\bibfnamefont {M.}~\bibnamefont {Skolnick}}, \bibinfo
  {author} {\bibfnamefont {T.}~\bibnamefont {Virgili}}, \bibinfo {author}
  {\bibfnamefont {S.}~\bibnamefont {Walker}}, \ and\ \bibinfo {author}
  {\bibfnamefont {D.}~\bibnamefont {Whittaker}},\ }\href {\doibase
  10.1038/25692} {\bibfield  {journal} {\bibinfo  {journal} {Nature}\ }\textbf
  {\bibinfo {volume} {395}},\ \bibinfo {pages} {53} (\bibinfo {year}
  {1998})}\BibitemShut {NoStop}%
\bibitem [{\citenamefont {K{\'{e}}na-Cohen}\ and\ \citenamefont
  {Forrest}(2010)}]{kena-cohen2010}%
  \BibitemOpen
  \bibfield  {author} {\bibinfo {author} {\bibfnamefont {S.}~\bibnamefont
  {K{\'{e}}na-Cohen}}\ and\ \bibinfo {author} {\bibfnamefont {S.~R.}\
  \bibnamefont {Forrest}},\ }\href {\doibase 10.1038/nphoton.2010.86}
  {\bibfield  {journal} {\bibinfo  {journal} {Nature Photonics}\ }\textbf
  {\bibinfo {volume} {4}},\ \bibinfo {pages} {371} (\bibinfo {year}
  {2010})}\BibitemShut {NoStop}%
\bibitem [{\citenamefont {Daskalakis}\ \emph {et~al.}(2014)\citenamefont
  {Daskalakis}, \citenamefont {Maier}, \citenamefont {Murray},\ and\
  \citenamefont {K{\'{e}}na-Cohen}}]{Daskalakis2014}%
  \BibitemOpen
  \bibfield  {author} {\bibinfo {author} {\bibfnamefont {K.~S.}\ \bibnamefont
  {Daskalakis}}, \bibinfo {author} {\bibfnamefont {S.~A.}\ \bibnamefont
  {Maier}}, \bibinfo {author} {\bibfnamefont {R.}~\bibnamefont {Murray}}, \
  and\ \bibinfo {author} {\bibfnamefont {S.}~\bibnamefont {K{\'{e}}na-Cohen}},\
  }\href {\doibase 10.1038/nmat3874} {\bibfield  {journal} {\bibinfo  {journal}
  {Nature Materials}\ }\textbf {\bibinfo {volume} {13}},\ \bibinfo {pages}
  {271} (\bibinfo {year} {2014})}\BibitemShut {NoStop}%
\bibitem [{\citenamefont {Plumhof}\ \emph {et~al.}(2013)\citenamefont
  {Plumhof}, \citenamefont {Stöferle}, \citenamefont {Mai}, \citenamefont
  {Scherf},\ and\ \citenamefont {Mahrt}}]{Plumhof2013}%
  \BibitemOpen
  \bibfield  {author} {\bibinfo {author} {\bibfnamefont {J.~D.}\ \bibnamefont
  {Plumhof}}, \bibinfo {author} {\bibfnamefont {T.}~\bibnamefont {Stöferle}},
  \bibinfo {author} {\bibfnamefont {L.}~\bibnamefont {Mai}}, \bibinfo {author}
  {\bibfnamefont {U.}~\bibnamefont {Scherf}}, \ and\ \bibinfo {author}
  {\bibfnamefont {R.~F.}\ \bibnamefont {Mahrt}},\ }\href {\doibase
  10.1038/nmat3825} {\bibfield  {journal} {\bibinfo  {journal} {Nature
  Materials}\ }\textbf {\bibinfo {volume} {13}},\ \bibinfo {pages} {247}
  (\bibinfo {year} {2013})}\BibitemShut {NoStop}%
\bibitem [{\citenamefont {Brehier}\ \emph {et~al.}(2006)\citenamefont
  {Brehier}, \citenamefont {Parashkov}, \citenamefont {Lauret},\ and\
  \citenamefont {Deleporte}}]{brehier2006strong}%
  \BibitemOpen
  \bibfield  {author} {\bibinfo {author} {\bibfnamefont {A.}~\bibnamefont
  {Brehier}}, \bibinfo {author} {\bibfnamefont {R.}~\bibnamefont {Parashkov}},
  \bibinfo {author} {\bibfnamefont {J.-S.}\ \bibnamefont {Lauret}}, \ and\
  \bibinfo {author} {\bibfnamefont {E.}~\bibnamefont {Deleporte}},\ }\href
  {\doibase 10.1063/1.2369533} {\bibfield  {journal} {\bibinfo  {journal}
  {Applied physics letters}\ }\textbf {\bibinfo {volume} {89}},\ \bibinfo
  {pages} {171110} (\bibinfo {year} {2006})}\BibitemShut {NoStop}%
\bibitem [{\citenamefont {Lanty}\ \emph {et~al.}(2008)\citenamefont {Lanty},
  \citenamefont {Brehier}, \citenamefont {Parashkov}, \citenamefont {Lauret},\
  and\ \citenamefont {Deleporte}}]{lanty2008strong}%
  \BibitemOpen
  \bibfield  {author} {\bibinfo {author} {\bibfnamefont {G.}~\bibnamefont
  {Lanty}}, \bibinfo {author} {\bibfnamefont {A.}~\bibnamefont {Brehier}},
  \bibinfo {author} {\bibfnamefont {R.}~\bibnamefont {Parashkov}}, \bibinfo
  {author} {\bibfnamefont {J.-S.}\ \bibnamefont {Lauret}}, \ and\ \bibinfo
  {author} {\bibfnamefont {E.}~\bibnamefont {Deleporte}},\ }\href {\doibase
  10.1088/1367-2630/10/6/065007} {\bibfield  {journal} {\bibinfo  {journal}
  {New Journal of Physics}\ }\textbf {\bibinfo {volume} {10}},\ \bibinfo
  {pages} {065007} (\bibinfo {year} {2008})}\BibitemShut {NoStop}%
\bibitem [{\citenamefont {Han}\ \emph {et~al.}(2012)\citenamefont {Han},
  \citenamefont {Nguyen}, \citenamefont {Boitier}, \citenamefont {Wei},
  \citenamefont {Abdel-Baki}, \citenamefont {Lauret}, \citenamefont {Bloch},
  \citenamefont {Bouchoule},\ and\ \citenamefont {Deleporte}}]{han2012high}%
  \BibitemOpen
  \bibfield  {author} {\bibinfo {author} {\bibfnamefont {Z.}~\bibnamefont
  {Han}}, \bibinfo {author} {\bibfnamefont {H.-S.}\ \bibnamefont {Nguyen}},
  \bibinfo {author} {\bibfnamefont {F.}~\bibnamefont {Boitier}}, \bibinfo
  {author} {\bibfnamefont {Y.}~\bibnamefont {Wei}}, \bibinfo {author}
  {\bibfnamefont {K.}~\bibnamefont {Abdel-Baki}}, \bibinfo {author}
  {\bibfnamefont {J.-S.}\ \bibnamefont {Lauret}}, \bibinfo {author}
  {\bibfnamefont {J.}~\bibnamefont {Bloch}}, \bibinfo {author} {\bibfnamefont
  {S.}~\bibnamefont {Bouchoule}}, \ and\ \bibinfo {author} {\bibfnamefont
  {E.}~\bibnamefont {Deleporte}},\ }\href {\doibase 10.1364/OL.37.005061}
  {\bibfield  {journal} {\bibinfo  {journal} {Optics letters}\ }\textbf
  {\bibinfo {volume} {37}},\ \bibinfo {pages} {5061} (\bibinfo {year}
  {2012})}\BibitemShut {NoStop}%
\bibitem [{\citenamefont {Fujita}\ \emph {et~al.}(1998)\citenamefont {Fujita},
  \citenamefont {Sato}, \citenamefont {Kuitani},\ and\ \citenamefont
  {Ishihara}}]{Fujita1998}%
  \BibitemOpen
  \bibfield  {author} {\bibinfo {author} {\bibfnamefont {T.}~\bibnamefont
  {Fujita}}, \bibinfo {author} {\bibfnamefont {Y.}~\bibnamefont {Sato}},
  \bibinfo {author} {\bibfnamefont {T.}~\bibnamefont {Kuitani}}, \ and\
  \bibinfo {author} {\bibfnamefont {T.}~\bibnamefont {Ishihara}},\ }\href
  {\doibase 10.1103/physrevb.57.12428} {\bibfield  {journal} {\bibinfo
  {journal} {Physical Review B}\ }\textbf {\bibinfo {volume} {57}},\ \bibinfo
  {pages} {12428} (\bibinfo {year} {1998})}\BibitemShut {NoStop}%
\bibitem [{\citenamefont {Nguyen}\ \emph {et~al.}(2014)\citenamefont {Nguyen},
  \citenamefont {Han}, \citenamefont {Abdel-Baki}, \citenamefont {Lafosse},
  \citenamefont {Amo}, \citenamefont {Lauret}, \citenamefont {Deleporte},
  \citenamefont {Bouchoule},\ and\ \citenamefont {Bloch}}]{nguyen2014quantum}%
  \BibitemOpen
  \bibfield  {author} {\bibinfo {author} {\bibfnamefont {H.~S.}\ \bibnamefont
  {Nguyen}}, \bibinfo {author} {\bibfnamefont {Z.}~\bibnamefont {Han}},
  \bibinfo {author} {\bibfnamefont {K.}~\bibnamefont {Abdel-Baki}}, \bibinfo
  {author} {\bibfnamefont {X.}~\bibnamefont {Lafosse}}, \bibinfo {author}
  {\bibfnamefont {A.}~\bibnamefont {Amo}}, \bibinfo {author} {\bibfnamefont
  {J.-S.}\ \bibnamefont {Lauret}}, \bibinfo {author} {\bibfnamefont
  {E.}~\bibnamefont {Deleporte}}, \bibinfo {author} {\bibfnamefont
  {S.}~\bibnamefont {Bouchoule}}, \ and\ \bibinfo {author} {\bibfnamefont
  {J.}~\bibnamefont {Bloch}},\ }\href {\doibase 10.1063/1.4866606} {\bibfield
  {journal} {\bibinfo  {journal} {Applied Physics Letters}\ }\textbf {\bibinfo
  {volume} {104}},\ \bibinfo {pages} {081103} (\bibinfo {year}
  {2014})}\BibitemShut {NoStop}%
\bibitem [{\citenamefont {Wang}\ \emph {et~al.}(2018)\citenamefont {Wang},
  \citenamefont {Su}, \citenamefont {Xing}, \citenamefont {Bao}, \citenamefont
  {Diederichs}, \citenamefont {Liu}, \citenamefont {Liew}, \citenamefont
  {Chen},\ and\ \citenamefont {Xiong}}]{Wang2018g}%
  \BibitemOpen
  \bibfield  {author} {\bibinfo {author} {\bibfnamefont {J.}~\bibnamefont
  {Wang}}, \bibinfo {author} {\bibfnamefont {R.}~\bibnamefont {Su}}, \bibinfo
  {author} {\bibfnamefont {J.}~\bibnamefont {Xing}}, \bibinfo {author}
  {\bibfnamefont {D.}~\bibnamefont {Bao}}, \bibinfo {author} {\bibfnamefont
  {C.}~\bibnamefont {Diederichs}}, \bibinfo {author} {\bibfnamefont
  {S.}~\bibnamefont {Liu}}, \bibinfo {author} {\bibfnamefont {T.~C.}\
  \bibnamefont {Liew}}, \bibinfo {author} {\bibfnamefont {Z.}~\bibnamefont
  {Chen}}, \ and\ \bibinfo {author} {\bibfnamefont {Q.}~\bibnamefont {Xiong}},\
  }\href {\doibase 10.1021/acsnano.8b03737} {\bibfield  {journal} {\bibinfo
  {journal} {{ACS} Nano}\ } (\bibinfo {year} {2018}),\
  10.1021/acsnano.8b03737}\BibitemShut {NoStop}%
\bibitem [{\citenamefont {Dufferwiel}\ \emph {et~al.}(2015)\citenamefont
  {Dufferwiel}, \citenamefont {Schwarz}, \citenamefont {Withers}, \citenamefont
  {Trichet}, \citenamefont {Li}, \citenamefont {Sich}, \citenamefont
  {Pozo-Zamudio}, \citenamefont {Clark}, \citenamefont {Nalitov}, \citenamefont
  {Solnyshkov}, \citenamefont {Malpuech}, \citenamefont {Novoselov},
  \citenamefont {Smith}, \citenamefont {Skolnick}, \citenamefont
  {Krizhanovskii},\ and\ \citenamefont {Tartakovskii}}]{Dufferwiel2015}%
  \BibitemOpen
  \bibfield  {author} {\bibinfo {author} {\bibfnamefont {S.}~\bibnamefont
  {Dufferwiel}}, \bibinfo {author} {\bibfnamefont {S.}~\bibnamefont {Schwarz}},
  \bibinfo {author} {\bibfnamefont {F.}~\bibnamefont {Withers}}, \bibinfo
  {author} {\bibfnamefont {A.~A.~P.}\ \bibnamefont {Trichet}}, \bibinfo
  {author} {\bibfnamefont {F.}~\bibnamefont {Li}}, \bibinfo {author}
  {\bibfnamefont {M.}~\bibnamefont {Sich}}, \bibinfo {author} {\bibfnamefont
  {O.~D.}\ \bibnamefont {Pozo-Zamudio}}, \bibinfo {author} {\bibfnamefont
  {C.}~\bibnamefont {Clark}}, \bibinfo {author} {\bibfnamefont
  {A.}~\bibnamefont {Nalitov}}, \bibinfo {author} {\bibfnamefont {D.~D.}\
  \bibnamefont {Solnyshkov}}, \bibinfo {author} {\bibfnamefont
  {G.}~\bibnamefont {Malpuech}}, \bibinfo {author} {\bibfnamefont {K.~S.}\
  \bibnamefont {Novoselov}}, \bibinfo {author} {\bibfnamefont {J.~M.}\
  \bibnamefont {Smith}}, \bibinfo {author} {\bibfnamefont {M.~S.}\ \bibnamefont
  {Skolnick}}, \bibinfo {author} {\bibfnamefont {D.~N.}\ \bibnamefont
  {Krizhanovskii}}, \ and\ \bibinfo {author} {\bibfnamefont {A.~I.}\
  \bibnamefont {Tartakovskii}},\ }\href {\doibase 10.1038/ncomms9579}
  {\bibfield  {journal} {\bibinfo  {journal} {Nature Communications}\ }\textbf
  {\bibinfo {volume} {6}} (\bibinfo {year} {2015}),\
  10.1038/ncomms9579}\BibitemShut {NoStop}%
\bibitem [{\citenamefont {Liu}\ \emph {et~al.}(2015)\citenamefont {Liu},
  \citenamefont {Galfsky}, \citenamefont {Sun}, \citenamefont {Xia},
  \citenamefont {chen Lin}, \citenamefont {Lee}, \citenamefont
  {K{\'{e}}na-Cohen},\ and\ \citenamefont {Menon}}]{Liu2015}%
  \BibitemOpen
  \bibfield  {author} {\bibinfo {author} {\bibfnamefont {X.}~\bibnamefont
  {Liu}}, \bibinfo {author} {\bibfnamefont {T.}~\bibnamefont {Galfsky}},
  \bibinfo {author} {\bibfnamefont {Z.}~\bibnamefont {Sun}}, \bibinfo {author}
  {\bibfnamefont {F.}~\bibnamefont {Xia}}, \bibinfo {author} {\bibfnamefont
  {E.}~\bibnamefont {chen Lin}}, \bibinfo {author} {\bibfnamefont {Y.-H.}\
  \bibnamefont {Lee}}, \bibinfo {author} {\bibfnamefont {S.}~\bibnamefont
  {K{\'{e}}na-Cohen}}, \ and\ \bibinfo {author} {\bibfnamefont {V.~M.}\
  \bibnamefont {Menon}},\ }\href {\doibase 10.1038/nphoton.2014.304} {\bibfield
   {journal} {\bibinfo  {journal} {Nature Photonics}\ }\textbf {\bibinfo
  {volume} {9}},\ \bibinfo {pages} {30} (\bibinfo {year} {2015})}\BibitemShut
  {NoStop}%
\bibitem [{NRE()}]{NREL}%
  \BibitemOpen
  \href@noop {} {\bibinfo  {journal} {NREL Efficiency Chart (accessed Oct1,
  2018) \\ https://www.nrel.gov/pv/assets/pdfs/pv-efficiencies-07-17-2018.pdf}\
  }\BibitemShut {NoStop}%
\bibitem [{\citenamefont {Kim}\ \emph {et~al.}(2014)\citenamefont {Kim},
  \citenamefont {Cho}, \citenamefont {Heo}, \citenamefont {Kim}, \citenamefont
  {Myoung}, \citenamefont {Lee}, \citenamefont {Im},\ and\ \citenamefont
  {Lee}}]{Kim2014}%
  \BibitemOpen
\bibfield  {journal} {  }\bibfield  {author} {\bibinfo {author} {\bibfnamefont
  {Y.-H.}\ \bibnamefont {Kim}}, \bibinfo {author} {\bibfnamefont
  {H.}~\bibnamefont {Cho}}, \bibinfo {author} {\bibfnamefont {J.~H.}\
  \bibnamefont {Heo}}, \bibinfo {author} {\bibfnamefont {T.-S.}\ \bibnamefont
  {Kim}}, \bibinfo {author} {\bibfnamefont {N.}~\bibnamefont {Myoung}},
  \bibinfo {author} {\bibfnamefont {C.-L.}\ \bibnamefont {Lee}}, \bibinfo
  {author} {\bibfnamefont {S.~H.}\ \bibnamefont {Im}}, \ and\ \bibinfo {author}
  {\bibfnamefont {T.-W.}\ \bibnamefont {Lee}},\ }\href {\doibase
  10.1002/adma.201403751} {\bibfield  {journal} {\bibinfo  {journal} {Advanced
  Materials}\ }\textbf {\bibinfo {volume} {27}},\ \bibinfo {pages} {1248}
  (\bibinfo {year} {2014})}\BibitemShut {NoStop}%
\bibitem [{\citenamefont {Tan}\ \emph {et~al.}(2014)\citenamefont {Tan},
  \citenamefont {Moghaddam}, \citenamefont {Lai}, \citenamefont {Docampo},
  \citenamefont {Higler}, \citenamefont {Deschler}, \citenamefont {Price},
  \citenamefont {Sadhanala}, \citenamefont {Pazos}, \citenamefont
  {Credgington}, \citenamefont {Hanusch}, \citenamefont {Bein}, \citenamefont
  {Snaith},\ and\ \citenamefont {Friend}}]{Tan2014}%
  \BibitemOpen
  \bibfield  {author} {\bibinfo {author} {\bibfnamefont {Z.-K.}\ \bibnamefont
  {Tan}}, \bibinfo {author} {\bibfnamefont {R.~S.}\ \bibnamefont {Moghaddam}},
  \bibinfo {author} {\bibfnamefont {M.~L.}\ \bibnamefont {Lai}}, \bibinfo
  {author} {\bibfnamefont {P.}~\bibnamefont {Docampo}}, \bibinfo {author}
  {\bibfnamefont {R.}~\bibnamefont {Higler}}, \bibinfo {author} {\bibfnamefont
  {F.}~\bibnamefont {Deschler}}, \bibinfo {author} {\bibfnamefont
  {M.}~\bibnamefont {Price}}, \bibinfo {author} {\bibfnamefont
  {A.}~\bibnamefont {Sadhanala}}, \bibinfo {author} {\bibfnamefont {L.~M.}\
  \bibnamefont {Pazos}}, \bibinfo {author} {\bibfnamefont {D.}~\bibnamefont
  {Credgington}}, \bibinfo {author} {\bibfnamefont {F.}~\bibnamefont
  {Hanusch}}, \bibinfo {author} {\bibfnamefont {T.}~\bibnamefont {Bein}},
  \bibinfo {author} {\bibfnamefont {H.~J.}\ \bibnamefont {Snaith}}, \ and\
  \bibinfo {author} {\bibfnamefont {R.~H.}\ \bibnamefont {Friend}},\ }\href
  {\doibase 10.1038/nnano.2014.149} {\bibfield  {journal} {\bibinfo  {journal}
  {Nature Nanotechnology}\ }\textbf {\bibinfo {volume} {9}},\ \bibinfo {pages}
  {687} (\bibinfo {year} {2014})}\BibitemShut {NoStop}%
\bibitem [{\citenamefont {Xing}\ \emph {et~al.}(2014)\citenamefont {Xing},
  \citenamefont {Mathews}, \citenamefont {Lim}, \citenamefont {Yantara},
  \citenamefont {Liu}, \citenamefont {Sabba}, \citenamefont {Grätzel},
  \citenamefont {Mhaisalkar},\ and\ \citenamefont {Sum}}]{Xing2014}%
  \BibitemOpen
  \bibfield  {author} {\bibinfo {author} {\bibfnamefont {G.}~\bibnamefont
  {Xing}}, \bibinfo {author} {\bibfnamefont {N.}~\bibnamefont {Mathews}},
  \bibinfo {author} {\bibfnamefont {S.~S.}\ \bibnamefont {Lim}}, \bibinfo
  {author} {\bibfnamefont {N.}~\bibnamefont {Yantara}}, \bibinfo {author}
  {\bibfnamefont {X.}~\bibnamefont {Liu}}, \bibinfo {author} {\bibfnamefont
  {D.}~\bibnamefont {Sabba}}, \bibinfo {author} {\bibfnamefont
  {M.}~\bibnamefont {Grätzel}}, \bibinfo {author} {\bibfnamefont
  {S.}~\bibnamefont {Mhaisalkar}}, \ and\ \bibinfo {author} {\bibfnamefont
  {T.~C.}\ \bibnamefont {Sum}},\ }\href {\doibase 10.1038/nmat3911} {\bibfield
  {journal} {\bibinfo  {journal} {Nature Materials}\ }\textbf {\bibinfo
  {volume} {13}},\ \bibinfo {pages} {476} (\bibinfo {year} {2014})}\BibitemShut
  {NoStop}%
\bibitem [{\citenamefont {Deschler}\ \emph {et~al.}(2014)\citenamefont
  {Deschler}, \citenamefont {Price}, \citenamefont {Pathak}, \citenamefont
  {Klintberg}, \citenamefont {Jarausch}, \citenamefont {Higler}, \citenamefont
  {H{\"u}ttner}, \citenamefont {Leijtens}, \citenamefont {Stranks},
  \citenamefont {Snaith} \emph {et~al.}}]{deschler2014high}%
  \BibitemOpen
  \bibfield  {author} {\bibinfo {author} {\bibfnamefont {F.}~\bibnamefont
  {Deschler}}, \bibinfo {author} {\bibfnamefont {M.}~\bibnamefont {Price}},
  \bibinfo {author} {\bibfnamefont {S.}~\bibnamefont {Pathak}}, \bibinfo
  {author} {\bibfnamefont {L.~E.}\ \bibnamefont {Klintberg}}, \bibinfo {author}
  {\bibfnamefont {D.-D.}\ \bibnamefont {Jarausch}}, \bibinfo {author}
  {\bibfnamefont {R.}~\bibnamefont {Higler}}, \bibinfo {author} {\bibfnamefont
  {S.}~\bibnamefont {H{\"u}ttner}}, \bibinfo {author} {\bibfnamefont
  {T.}~\bibnamefont {Leijtens}}, \bibinfo {author} {\bibfnamefont {S.~D.}\
  \bibnamefont {Stranks}}, \bibinfo {author} {\bibfnamefont {H.~J.}\
  \bibnamefont {Snaith}},  \emph {et~al.},\ }\href {\doibase 10.1021/jz5005285}
  {\bibfield  {journal} {\bibinfo  {journal} {The journal of physical chemistry
  letters}\ }\textbf {\bibinfo {volume} {5}},\ \bibinfo {pages} {1421}
  (\bibinfo {year} {2014})}\BibitemShut {NoStop}%
\bibitem [{\citenamefont {Zhu}\ \emph {et~al.}(2015)\citenamefont {Zhu},
  \citenamefont {Fu}, \citenamefont {Meng}, \citenamefont {Wu}, \citenamefont
  {Gong}, \citenamefont {Ding}, \citenamefont {Gustafsson}, \citenamefont
  {Trinh}, \citenamefont {Jin},\ and\ \citenamefont {Zhu}}]{zhu2015lead}%
  \BibitemOpen
  \bibfield  {author} {\bibinfo {author} {\bibfnamefont {H.}~\bibnamefont
  {Zhu}}, \bibinfo {author} {\bibfnamefont {Y.}~\bibnamefont {Fu}}, \bibinfo
  {author} {\bibfnamefont {F.}~\bibnamefont {Meng}}, \bibinfo {author}
  {\bibfnamefont {X.}~\bibnamefont {Wu}}, \bibinfo {author} {\bibfnamefont
  {Z.}~\bibnamefont {Gong}}, \bibinfo {author} {\bibfnamefont {Q.}~\bibnamefont
  {Ding}}, \bibinfo {author} {\bibfnamefont {M.~V.}\ \bibnamefont
  {Gustafsson}}, \bibinfo {author} {\bibfnamefont {M.~T.}\ \bibnamefont
  {Trinh}}, \bibinfo {author} {\bibfnamefont {S.}~\bibnamefont {Jin}}, \ and\
  \bibinfo {author} {\bibfnamefont {X.}~\bibnamefont {Zhu}},\ }\href {\doibase
  10.1038/NMAT4271} {\bibfield  {journal} {\bibinfo  {journal} {Nature
  materials}\ }\textbf {\bibinfo {volume} {14}},\ \bibinfo {pages} {636}
  (\bibinfo {year} {2015})}\BibitemShut {NoStop}%
\bibitem [{\citenamefont {Chen}\ and\ \citenamefont
  {Nurmikko}(2018)}]{Chen2018c}%
  \BibitemOpen
  \bibfield  {author} {\bibinfo {author} {\bibfnamefont {S.}~\bibnamefont
  {Chen}}\ and\ \bibinfo {author} {\bibfnamefont {A.}~\bibnamefont
  {Nurmikko}},\ }\href {\doibase 10.1364/optica.5.001141} {\bibfield  {journal}
  {\bibinfo  {journal} {Optica}\ }\textbf {\bibinfo {volume} {5}},\ \bibinfo
  {pages} {1141} (\bibinfo {year} {2018})}\BibitemShut {NoStop}%
\bibitem [{\citenamefont {Zhang}\ \emph {et~al.}(2016)\citenamefont {Zhang},
  \citenamefont {Su}, \citenamefont {Liu}, \citenamefont {Xing}, \citenamefont
  {Sum},\ and\ \citenamefont {Xiong}}]{Zhang2016b}%
  \BibitemOpen
  \bibfield  {author} {\bibinfo {author} {\bibfnamefont {Q.}~\bibnamefont
  {Zhang}}, \bibinfo {author} {\bibfnamefont {R.}~\bibnamefont {Su}}, \bibinfo
  {author} {\bibfnamefont {X.}~\bibnamefont {Liu}}, \bibinfo {author}
  {\bibfnamefont {J.}~\bibnamefont {Xing}}, \bibinfo {author} {\bibfnamefont
  {T.~C.}\ \bibnamefont {Sum}}, \ and\ \bibinfo {author} {\bibfnamefont
  {Q.}~\bibnamefont {Xiong}},\ }\href {\doibase 10.1002/adfm.201601690}
  {\bibfield  {journal} {\bibinfo  {journal} {Advanced Functional Materials}\
  }\textbf {\bibinfo {volume} {26}},\ \bibinfo {pages} {6238} (\bibinfo {year}
  {2016})}\BibitemShut {NoStop}%
\bibitem [{\citenamefont {Ponseca}\ \emph {et~al.}(2014)\citenamefont
  {Ponseca}, \citenamefont {Savenije}, \citenamefont {Abdellah}, \citenamefont
  {Zheng}, \citenamefont {Yartsev}, \citenamefont {Pascher}, \citenamefont
  {Harlang}, \citenamefont {Chabera}, \citenamefont {Pullerits}, \citenamefont
  {Stepanov}, \citenamefont {Wolf},\ and\ \citenamefont
  {Sundström}}]{Ponseca2014}%
  \BibitemOpen
  \bibfield  {author} {\bibinfo {author} {\bibfnamefont {C.~S.}\ \bibnamefont
  {Ponseca}}, \bibinfo {author} {\bibfnamefont {T.~J.}\ \bibnamefont
  {Savenije}}, \bibinfo {author} {\bibfnamefont {M.}~\bibnamefont {Abdellah}},
  \bibinfo {author} {\bibfnamefont {K.}~\bibnamefont {Zheng}}, \bibinfo
  {author} {\bibfnamefont {A.}~\bibnamefont {Yartsev}}, \bibinfo {author}
  {\bibfnamefont {T.}~\bibnamefont {Pascher}}, \bibinfo {author} {\bibfnamefont
  {T.}~\bibnamefont {Harlang}}, \bibinfo {author} {\bibfnamefont
  {P.}~\bibnamefont {Chabera}}, \bibinfo {author} {\bibfnamefont
  {T.}~\bibnamefont {Pullerits}}, \bibinfo {author} {\bibfnamefont
  {A.}~\bibnamefont {Stepanov}}, \bibinfo {author} {\bibfnamefont {J.-P.}\
  \bibnamefont {Wolf}}, \ and\ \bibinfo {author} {\bibfnamefont
  {V.}~\bibnamefont {Sundström}},\ }\href {\doibase 10.1021/ja412583t}
  {\bibfield  {journal} {\bibinfo  {journal} {Journal of the American Chemical
  Society}\ }\textbf {\bibinfo {volume} {136}},\ \bibinfo {pages} {5189}
  (\bibinfo {year} {2014})}\BibitemShut {NoStop}%
\bibitem [{\citenamefont {Xing}\ \emph {et~al.}(2013)\citenamefont {Xing},
  \citenamefont {Mathews}, \citenamefont {Sun}, \citenamefont {Lim},
  \citenamefont {Lam}, \citenamefont {Gratzel}, \citenamefont {Mhaisalkar},\
  and\ \citenamefont {Sum}}]{Xing2013}%
  \BibitemOpen
  \bibfield  {author} {\bibinfo {author} {\bibfnamefont {G.}~\bibnamefont
  {Xing}}, \bibinfo {author} {\bibfnamefont {N.}~\bibnamefont {Mathews}},
  \bibinfo {author} {\bibfnamefont {S.}~\bibnamefont {Sun}}, \bibinfo {author}
  {\bibfnamefont {S.~S.}\ \bibnamefont {Lim}}, \bibinfo {author} {\bibfnamefont
  {Y.~M.}\ \bibnamefont {Lam}}, \bibinfo {author} {\bibfnamefont
  {M.}~\bibnamefont {Gratzel}}, \bibinfo {author} {\bibfnamefont
  {S.}~\bibnamefont {Mhaisalkar}}, \ and\ \bibinfo {author} {\bibfnamefont
  {T.~C.}\ \bibnamefont {Sum}},\ }\href {\doibase 10.1126/science.1243167}
  {\bibfield  {journal} {\bibinfo  {journal} {Science}\ }\textbf {\bibinfo
  {volume} {342}},\ \bibinfo {pages} {344} (\bibinfo {year}
  {2013})}\BibitemShut {NoStop}%
\bibitem [{\citenamefont {Miyata}\ \emph {et~al.}(2015)\citenamefont {Miyata},
  \citenamefont {Mitioglu}, \citenamefont {Plochocka}, \citenamefont
  {Portugall}, \citenamefont {Wang}, \citenamefont {Stranks}, \citenamefont
  {Snaith},\ and\ \citenamefont {Nicholas}}]{miyata2015direct}%
  \BibitemOpen
  \bibfield  {author} {\bibinfo {author} {\bibfnamefont {A.}~\bibnamefont
  {Miyata}}, \bibinfo {author} {\bibfnamefont {A.}~\bibnamefont {Mitioglu}},
  \bibinfo {author} {\bibfnamefont {P.}~\bibnamefont {Plochocka}}, \bibinfo
  {author} {\bibfnamefont {O.}~\bibnamefont {Portugall}}, \bibinfo {author}
  {\bibfnamefont {J.~T.-W.}\ \bibnamefont {Wang}}, \bibinfo {author}
  {\bibfnamefont {S.~D.}\ \bibnamefont {Stranks}}, \bibinfo {author}
  {\bibfnamefont {H.~J.}\ \bibnamefont {Snaith}}, \ and\ \bibinfo {author}
  {\bibfnamefont {R.~J.}\ \bibnamefont {Nicholas}},\ }\href {\doibase
  10.1038/NPHYS3357} {\bibfield  {journal} {\bibinfo  {journal} {Nature
  Physics}\ }\textbf {\bibinfo {volume} {11}},\ \bibinfo {pages} {582}
  (\bibinfo {year} {2015})}\BibitemShut {NoStop}%
\bibitem [{\citenamefont {Phuong}\ \emph {et~al.}(2016)\citenamefont {Phuong},
  \citenamefont {Yamada}, \citenamefont {Nagai}, \citenamefont {Maruyama},
  \citenamefont {Wakamiya},\ and\ \citenamefont {Kanemitsu}}]{Phuong2016}%
  \BibitemOpen
  \bibfield  {author} {\bibinfo {author} {\bibfnamefont {L.~Q.}\ \bibnamefont
  {Phuong}}, \bibinfo {author} {\bibfnamefont {Y.}~\bibnamefont {Yamada}},
  \bibinfo {author} {\bibfnamefont {M.}~\bibnamefont {Nagai}}, \bibinfo
  {author} {\bibfnamefont {N.}~\bibnamefont {Maruyama}}, \bibinfo {author}
  {\bibfnamefont {A.}~\bibnamefont {Wakamiya}}, \ and\ \bibinfo {author}
  {\bibfnamefont {Y.}~\bibnamefont {Kanemitsu}},\ }\href {\doibase
  10.1021/acs.jpclett.6b00781} {\bibfield  {journal} {\bibinfo  {journal} {The
  Journal of Physical Chemistry Letters}\ }\textbf {\bibinfo {volume} {7}},\
  \bibinfo {pages} {2316} (\bibinfo {year} {2016})}\BibitemShut {NoStop}%
\bibitem [{\citenamefont {Yang}\ \emph {et~al.}(2017)\citenamefont {Yang},
  \citenamefont {Surrente}, \citenamefont {Galkowski}, \citenamefont {Bruyant},
  \citenamefont {Maude}, \citenamefont {Haghighirad}, \citenamefont {Snaith},
  \citenamefont {Plochocka},\ and\ \citenamefont {Nicholas}}]{Yang2017a}%
  \BibitemOpen
  \bibfield  {author} {\bibinfo {author} {\bibfnamefont {Z.}~\bibnamefont
  {Yang}}, \bibinfo {author} {\bibfnamefont {A.}~\bibnamefont {Surrente}},
  \bibinfo {author} {\bibfnamefont {K.}~\bibnamefont {Galkowski}}, \bibinfo
  {author} {\bibfnamefont {N.}~\bibnamefont {Bruyant}}, \bibinfo {author}
  {\bibfnamefont {D.~K.}\ \bibnamefont {Maude}}, \bibinfo {author}
  {\bibfnamefont {A.~A.}\ \bibnamefont {Haghighirad}}, \bibinfo {author}
  {\bibfnamefont {H.~J.}\ \bibnamefont {Snaith}}, \bibinfo {author}
  {\bibfnamefont {P.}~\bibnamefont {Plochocka}}, \ and\ \bibinfo {author}
  {\bibfnamefont {R.~J.}\ \bibnamefont {Nicholas}},\ }\href {\doibase
  10.1021/acs.jpclett.7b00524} {\bibfield  {journal} {\bibinfo  {journal} {The
  Journal of Physical Chemistry Letters}\ }\textbf {\bibinfo {volume} {8}},\
  \bibinfo {pages} {1851} (\bibinfo {year} {2017})}\BibitemShut {NoStop}%
\bibitem [{\citenamefont {Comin}\ \emph {et~al.}(2015)\citenamefont {Comin},
  \citenamefont {Walters}, \citenamefont {Thibau}, \citenamefont {Voznyy},
  \citenamefont {Lu},\ and\ \citenamefont {Sargent}}]{Comin2015}%
  \BibitemOpen
  \bibfield  {author} {\bibinfo {author} {\bibfnamefont {R.}~\bibnamefont
  {Comin}}, \bibinfo {author} {\bibfnamefont {G.}~\bibnamefont {Walters}},
  \bibinfo {author} {\bibfnamefont {E.~S.}\ \bibnamefont {Thibau}}, \bibinfo
  {author} {\bibfnamefont {O.}~\bibnamefont {Voznyy}}, \bibinfo {author}
  {\bibfnamefont {Z.-H.}\ \bibnamefont {Lu}}, \ and\ \bibinfo {author}
  {\bibfnamefont {E.~H.}\ \bibnamefont {Sargent}},\ }\href {\doibase
  10.1039/c5tc01718a} {\bibfield  {journal} {\bibinfo  {journal} {Journal of
  Materials Chemistry C}\ }\textbf {\bibinfo {volume} {3}},\ \bibinfo {pages}
  {8839} (\bibinfo {year} {2015})}\BibitemShut {NoStop}%
\bibitem [{\citenamefont {Saba}\ \emph {et~al.}(2015)\citenamefont {Saba},
  \citenamefont {Quochi}, \citenamefont {Mura},\ and\ \citenamefont
  {Bongiovanni}}]{Saba2015}%
  \BibitemOpen
  \bibfield  {author} {\bibinfo {author} {\bibfnamefont {M.}~\bibnamefont
  {Saba}}, \bibinfo {author} {\bibfnamefont {F.}~\bibnamefont {Quochi}},
  \bibinfo {author} {\bibfnamefont {A.}~\bibnamefont {Mura}}, \ and\ \bibinfo
  {author} {\bibfnamefont {G.}~\bibnamefont {Bongiovanni}},\ }\href {\doibase
  10.1021/acs.accounts.5b00445} {\bibfield  {journal} {\bibinfo  {journal}
  {Accounts of Chemical Research}\ }\textbf {\bibinfo {volume} {49}},\ \bibinfo
  {pages} {166} (\bibinfo {year} {2015})}\BibitemShut {NoStop}%
\bibitem [{\citenamefont {Yamada}\ \emph {et~al.}(2018)\citenamefont {Yamada},
  \citenamefont {Aharen},\ and\ \citenamefont {Kanemitsu}}]{Yamada2018}%
  \BibitemOpen
  \bibfield  {author} {\bibinfo {author} {\bibfnamefont {T.}~\bibnamefont
  {Yamada}}, \bibinfo {author} {\bibfnamefont {T.}~\bibnamefont {Aharen}}, \
  and\ \bibinfo {author} {\bibfnamefont {Y.}~\bibnamefont {Kanemitsu}},\ }\href
  {\doibase 10.1103/physrevlett.120.057404} {\bibfield  {journal} {\bibinfo
  {journal} {Physical Review Letters}\ }\textbf {\bibinfo {volume} {120}}
  (\bibinfo {year} {2018}),\ 10.1103/physrevlett.120.057404}\BibitemShut
  {NoStop}%
\bibitem [{\citenamefont {Protesescu}\ \emph {et~al.}(2015)\citenamefont
  {Protesescu}, \citenamefont {Yakunin}, \citenamefont {Bodnarchuk},
  \citenamefont {Krieg}, \citenamefont {Caputo}, \citenamefont {Hendon},
  \citenamefont {Yang}, \citenamefont {Walsh},\ and\ \citenamefont
  {Kovalenko}}]{Protesescu2015}%
  \BibitemOpen
  \bibfield  {author} {\bibinfo {author} {\bibfnamefont {L.}~\bibnamefont
  {Protesescu}}, \bibinfo {author} {\bibfnamefont {S.}~\bibnamefont {Yakunin}},
  \bibinfo {author} {\bibfnamefont {M.~I.}\ \bibnamefont {Bodnarchuk}},
  \bibinfo {author} {\bibfnamefont {F.}~\bibnamefont {Krieg}}, \bibinfo
  {author} {\bibfnamefont {R.}~\bibnamefont {Caputo}}, \bibinfo {author}
  {\bibfnamefont {C.~H.}\ \bibnamefont {Hendon}}, \bibinfo {author}
  {\bibfnamefont {R.~X.}\ \bibnamefont {Yang}}, \bibinfo {author}
  {\bibfnamefont {A.}~\bibnamefont {Walsh}}, \ and\ \bibinfo {author}
  {\bibfnamefont {M.~V.}\ \bibnamefont {Kovalenko}},\ }\href {\doibase
  10.1021/nl5048779} {\bibfield  {journal} {\bibinfo  {journal} {Nano Letters}\
  }\textbf {\bibinfo {volume} {15}},\ \bibinfo {pages} {3692} (\bibinfo {year}
  {2015})}\BibitemShut {NoStop}%
\bibitem [{\citenamefont {Su}\ \emph {et~al.}(2017)\citenamefont {Su},
  \citenamefont {Diederichs}, \citenamefont {Wang}, \citenamefont {Liew},
  \citenamefont {Zhao}, \citenamefont {Liu}, \citenamefont {Xu}, \citenamefont
  {Chen},\ and\ \citenamefont {Xiong}}]{su2017room}%
  \BibitemOpen
  \bibfield  {author} {\bibinfo {author} {\bibfnamefont {R.}~\bibnamefont
  {Su}}, \bibinfo {author} {\bibfnamefont {C.}~\bibnamefont {Diederichs}},
  \bibinfo {author} {\bibfnamefont {J.}~\bibnamefont {Wang}}, \bibinfo {author}
  {\bibfnamefont {T.~C.}\ \bibnamefont {Liew}}, \bibinfo {author}
  {\bibfnamefont {J.}~\bibnamefont {Zhao}}, \bibinfo {author} {\bibfnamefont
  {S.}~\bibnamefont {Liu}}, \bibinfo {author} {\bibfnamefont {W.}~\bibnamefont
  {Xu}}, \bibinfo {author} {\bibfnamefont {Z.}~\bibnamefont {Chen}}, \ and\
  \bibinfo {author} {\bibfnamefont {Q.}~\bibnamefont {Xiong}},\ }\href
  {\doibase 10.1021/acs.nanolett.7b01956} {\bibfield  {journal} {\bibinfo
  {journal} {Nano letters}\ }\textbf {\bibinfo {volume} {17}},\ \bibinfo
  {pages} {3982} (\bibinfo {year} {2017})}\BibitemShut {NoStop}%
\bibitem [{\citenamefont {Tanaka}\ \emph {et~al.}(2003)\citenamefont {Tanaka},
  \citenamefont {Takahashi}, \citenamefont {Ban}, \citenamefont {Kondo},
  \citenamefont {Uchida},\ and\ \citenamefont {Miura}}]{tanaka2003comparative}%
  \BibitemOpen
  \bibfield  {author} {\bibinfo {author} {\bibfnamefont {K.}~\bibnamefont
  {Tanaka}}, \bibinfo {author} {\bibfnamefont {T.}~\bibnamefont {Takahashi}},
  \bibinfo {author} {\bibfnamefont {T.}~\bibnamefont {Ban}}, \bibinfo {author}
  {\bibfnamefont {T.}~\bibnamefont {Kondo}}, \bibinfo {author} {\bibfnamefont
  {K.}~\bibnamefont {Uchida}}, \ and\ \bibinfo {author} {\bibfnamefont
  {N.}~\bibnamefont {Miura}},\ }\href {\doibase 10.1016/S0038-1098(03)00566-0}
  {\bibfield  {journal} {\bibinfo  {journal} {Solid state communications}\
  }\textbf {\bibinfo {volume} {127}},\ \bibinfo {pages} {619} (\bibinfo {year}
  {2003})}\BibitemShut {NoStop}%
\bibitem [{\citenamefont {Yang}\ \emph {et~al.}(2015)\citenamefont {Yang},
  \citenamefont {Yang}, \citenamefont {Li}, \citenamefont {Crisp},
  \citenamefont {Zhu},\ and\ \citenamefont {Beard}}]{Yang2015}%
  \BibitemOpen
  \bibfield  {author} {\bibinfo {author} {\bibfnamefont {Y.}~\bibnamefont
  {Yang}}, \bibinfo {author} {\bibfnamefont {M.}~\bibnamefont {Yang}}, \bibinfo
  {author} {\bibfnamefont {Z.}~\bibnamefont {Li}}, \bibinfo {author}
  {\bibfnamefont {R.}~\bibnamefont {Crisp}}, \bibinfo {author} {\bibfnamefont
  {K.}~\bibnamefont {Zhu}}, \ and\ \bibinfo {author} {\bibfnamefont {M.~C.}\
  \bibnamefont {Beard}},\ }\href {\doibase 10.1021/acs.jpclett.5b02290}
  {\bibfield  {journal} {\bibinfo  {journal} {The Journal of Physical Chemistry
  Letters}\ }\textbf {\bibinfo {volume} {6}},\ \bibinfo {pages} {4688}
  (\bibinfo {year} {2015})}\BibitemShut {NoStop}%
\bibitem [{\citenamefont {Soufiani}\ \emph {et~al.}(2015)\citenamefont
  {Soufiani}, \citenamefont {Huang}, \citenamefont {Reece}, \citenamefont
  {Sheng}, \citenamefont {Ho-Baillie},\ and\ \citenamefont
  {Green}}]{soufiani2015polaronic}%
  \BibitemOpen
  \bibfield  {author} {\bibinfo {author} {\bibfnamefont {A.~M.}\ \bibnamefont
  {Soufiani}}, \bibinfo {author} {\bibfnamefont {F.}~\bibnamefont {Huang}},
  \bibinfo {author} {\bibfnamefont {P.}~\bibnamefont {Reece}}, \bibinfo
  {author} {\bibfnamefont {R.}~\bibnamefont {Sheng}}, \bibinfo {author}
  {\bibfnamefont {A.}~\bibnamefont {Ho-Baillie}}, \ and\ \bibinfo {author}
  {\bibfnamefont {M.~A.}\ \bibnamefont {Green}},\ }\href {\doibase
  10.1063/1.4936418} {\bibfield  {journal} {\bibinfo  {journal} {Applied
  Physics Letters}\ }\textbf {\bibinfo {volume} {107}},\ \bibinfo {pages}
  {231902} (\bibinfo {year} {2015})}\BibitemShut {NoStop}%
\bibitem [{\citenamefont {Galkowski}\ \emph {et~al.}(2016)\citenamefont
  {Galkowski}, \citenamefont {Mitioglu}, \citenamefont {Miyata}, \citenamefont
  {Plochocka}, \citenamefont {Portugall}, \citenamefont {Eperon}, \citenamefont
  {Wang}, \citenamefont {Stergiopoulos}, \citenamefont {Stranks}, \citenamefont
  {Snaith},\ and\ \citenamefont {Nicholas}}]{Galkowski2016}%
  \BibitemOpen
  \bibfield  {author} {\bibinfo {author} {\bibfnamefont {K.}~\bibnamefont
  {Galkowski}}, \bibinfo {author} {\bibfnamefont {A.}~\bibnamefont {Mitioglu}},
  \bibinfo {author} {\bibfnamefont {A.}~\bibnamefont {Miyata}}, \bibinfo
  {author} {\bibfnamefont {P.}~\bibnamefont {Plochocka}}, \bibinfo {author}
  {\bibfnamefont {O.}~\bibnamefont {Portugall}}, \bibinfo {author}
  {\bibfnamefont {G.~E.}\ \bibnamefont {Eperon}}, \bibinfo {author}
  {\bibfnamefont {J.~T.-W.}\ \bibnamefont {Wang}}, \bibinfo {author}
  {\bibfnamefont {T.}~\bibnamefont {Stergiopoulos}}, \bibinfo {author}
  {\bibfnamefont {S.~D.}\ \bibnamefont {Stranks}}, \bibinfo {author}
  {\bibfnamefont {H.~J.}\ \bibnamefont {Snaith}}, \ and\ \bibinfo {author}
  {\bibfnamefont {R.~J.}\ \bibnamefont {Nicholas}},\ }\href {\doibase
  10.1039/c5ee03435c} {\bibfield  {journal} {\bibinfo  {journal} {Energy {\&}
  Environmental Science}\ }\textbf {\bibinfo {volume} {9}},\ \bibinfo {pages}
  {962} (\bibinfo {year} {2016})}\BibitemShut {NoStop}%
\bibitem [{\citenamefont {Tilchin}\ \emph {et~al.}(2016)\citenamefont
  {Tilchin}, \citenamefont {Dirin}, \citenamefont {Maikov}, \citenamefont
  {Sashchiuk}, \citenamefont {Kovalenko},\ and\ \citenamefont
  {Lifshitz}}]{tilchin2016hydrogen}%
  \BibitemOpen
  \bibfield  {author} {\bibinfo {author} {\bibfnamefont {J.}~\bibnamefont
  {Tilchin}}, \bibinfo {author} {\bibfnamefont {D.~N.}\ \bibnamefont {Dirin}},
  \bibinfo {author} {\bibfnamefont {G.~I.}\ \bibnamefont {Maikov}}, \bibinfo
  {author} {\bibfnamefont {A.}~\bibnamefont {Sashchiuk}}, \bibinfo {author}
  {\bibfnamefont {M.~V.}\ \bibnamefont {Kovalenko}}, \ and\ \bibinfo {author}
  {\bibfnamefont {E.}~\bibnamefont {Lifshitz}},\ }\href {\doibase
  10.1021/acsnano.6b02734} {\bibfield  {journal} {\bibinfo  {journal} {ACS
  nano}\ }\textbf {\bibinfo {volume} {10}},\ \bibinfo {pages} {6363} (\bibinfo
  {year} {2016})}\BibitemShut {NoStop}%
\bibitem [{\citenamefont {Wu}\ \emph {et~al.}(2016)\citenamefont {Wu},
  \citenamefont {Nguyen}, \citenamefont {Ku}, \citenamefont {Han},
  \citenamefont {Giovanni}, \citenamefont {Mathews}, \citenamefont {Fan},\ and\
  \citenamefont {Sum}}]{Wu2016}%
  \BibitemOpen
  \bibfield  {author} {\bibinfo {author} {\bibfnamefont {B.}~\bibnamefont
  {Wu}}, \bibinfo {author} {\bibfnamefont {H.~T.}\ \bibnamefont {Nguyen}},
  \bibinfo {author} {\bibfnamefont {Z.}~\bibnamefont {Ku}}, \bibinfo {author}
  {\bibfnamefont {G.}~\bibnamefont {Han}}, \bibinfo {author} {\bibfnamefont
  {D.}~\bibnamefont {Giovanni}}, \bibinfo {author} {\bibfnamefont
  {N.}~\bibnamefont {Mathews}}, \bibinfo {author} {\bibfnamefont {H.~J.}\
  \bibnamefont {Fan}}, \ and\ \bibinfo {author} {\bibfnamefont {T.~C.}\
  \bibnamefont {Sum}},\ }\href {\doibase 10.1002/aenm.201600551} {\bibfield
  {journal} {\bibinfo  {journal} {Advanced Energy Materials}\ }\textbf
  {\bibinfo {volume} {6}},\ \bibinfo {pages} {1600551} (\bibinfo {year}
  {2016})}\BibitemShut {NoStop}%
\bibitem [{\citenamefont {Wolf}\ \emph {et~al.}(2017)\citenamefont {Wolf},
  \citenamefont {Kim},\ and\ \citenamefont {Lee}}]{wolf2017structural}%
  \BibitemOpen
  \bibfield  {author} {\bibinfo {author} {\bibfnamefont {C.}~\bibnamefont
  {Wolf}}, \bibinfo {author} {\bibfnamefont {J.-S.}\ \bibnamefont {Kim}}, \
  and\ \bibinfo {author} {\bibfnamefont {T.-W.}\ \bibnamefont {Lee}},\ }\href
  {\doibase 10.1021/acsami.6b15694} {\bibfield  {journal} {\bibinfo  {journal}
  {ACS applied materials \& interfaces}\ }\textbf {\bibinfo {volume} {9}},\
  \bibinfo {pages} {10344} (\bibinfo {year} {2017})}\BibitemShut {NoStop}%
\bibitem [{\citenamefont {Niesner}\ \emph {et~al.}(2017)\citenamefont
  {Niesner}, \citenamefont {Schuster}, \citenamefont {Wilhelm}, \citenamefont
  {Levchuk}, \citenamefont {Osvet}, \citenamefont {Shrestha}, \citenamefont
  {Batentschuk}, \citenamefont {Brabec},\ and\ \citenamefont
  {Fauster}}]{Niesner2017}%
  \BibitemOpen
  \bibfield  {author} {\bibinfo {author} {\bibfnamefont {D.}~\bibnamefont
  {Niesner}}, \bibinfo {author} {\bibfnamefont {O.}~\bibnamefont {Schuster}},
  \bibinfo {author} {\bibfnamefont {M.}~\bibnamefont {Wilhelm}}, \bibinfo
  {author} {\bibfnamefont {I.}~\bibnamefont {Levchuk}}, \bibinfo {author}
  {\bibfnamefont {A.}~\bibnamefont {Osvet}}, \bibinfo {author} {\bibfnamefont
  {S.}~\bibnamefont {Shrestha}}, \bibinfo {author} {\bibfnamefont
  {M.}~\bibnamefont {Batentschuk}}, \bibinfo {author} {\bibfnamefont
  {C.}~\bibnamefont {Brabec}}, \ and\ \bibinfo {author} {\bibfnamefont
  {T.}~\bibnamefont {Fauster}},\ }\href {\doibase 10.1103/physrevb.95.075207}
  {\bibfield  {journal} {\bibinfo  {journal} {Physical Review B}\ }\textbf
  {\bibinfo {volume} {95}} (\bibinfo {year} {2017}),\
  10.1103/physrevb.95.075207}\BibitemShut {NoStop}%
\bibitem [{\citenamefont {Droseros}\ \emph {et~al.}(2018)\citenamefont
  {Droseros}, \citenamefont {Longo}, \citenamefont {Brauer}, \citenamefont
  {Sessolo}, \citenamefont {Bolink},\ and\ \citenamefont
  {Banerji}}]{Droseros2018}%
  \BibitemOpen
  \bibfield  {author} {\bibinfo {author} {\bibfnamefont {N.}~\bibnamefont
  {Droseros}}, \bibinfo {author} {\bibfnamefont {G.}~\bibnamefont {Longo}},
  \bibinfo {author} {\bibfnamefont {J.~C.}\ \bibnamefont {Brauer}}, \bibinfo
  {author} {\bibfnamefont {M.}~\bibnamefont {Sessolo}}, \bibinfo {author}
  {\bibfnamefont {H.~J.}\ \bibnamefont {Bolink}}, \ and\ \bibinfo {author}
  {\bibfnamefont {N.}~\bibnamefont {Banerji}},\ }\href {\doibase
  10.1021/acsenergylett.8b00475} {\bibfield  {journal} {\bibinfo  {journal}
  {{ACS} Energy Letters}\ }\textbf {\bibinfo {volume} {3}},\ \bibinfo {pages}
  {1458} (\bibinfo {year} {2018})}\BibitemShut {NoStop}%
\bibitem [{\citenamefont {Park}\ \emph {et~al.}(2016)\citenamefont {Park},
  \citenamefont {Lee}, \citenamefont {Kim}, \citenamefont {Han}, \citenamefont
  {Jang}, \citenamefont {Jeong}, \citenamefont {Park},\ and\ \citenamefont
  {Song}}]{Park2016}%
  \BibitemOpen
  \bibfield  {author} {\bibinfo {author} {\bibfnamefont {K.}~\bibnamefont
  {Park}}, \bibinfo {author} {\bibfnamefont {J.~W.}\ \bibnamefont {Lee}},
  \bibinfo {author} {\bibfnamefont {J.~D.}\ \bibnamefont {Kim}}, \bibinfo
  {author} {\bibfnamefont {N.~S.}\ \bibnamefont {Han}}, \bibinfo {author}
  {\bibfnamefont {D.~M.}\ \bibnamefont {Jang}}, \bibinfo {author}
  {\bibfnamefont {S.}~\bibnamefont {Jeong}}, \bibinfo {author} {\bibfnamefont
  {J.}~\bibnamefont {Park}}, \ and\ \bibinfo {author} {\bibfnamefont {J.~K.}\
  \bibnamefont {Song}},\ }\href {\doibase 10.1021/acs.jpclett.6b01821}
  {\bibfield  {journal} {\bibinfo  {journal} {The Journal of Physical Chemistry
  Letters}\ }\textbf {\bibinfo {volume} {7}},\ \bibinfo {pages} {3703}
  (\bibinfo {year} {2016})}\BibitemShut {NoStop}%
\bibitem [{\citenamefont {Evans}\ \emph {et~al.}(2018)\citenamefont {Evans},
  \citenamefont {Schlaus}, \citenamefont {Fu}, \citenamefont {Zhong},
  \citenamefont {Atallah}, \citenamefont {Spencer}, \citenamefont {Brus},
  \citenamefont {Jin},\ and\ \citenamefont {Zhu}}]{evans2018continuous}%
  \BibitemOpen
  \bibfield  {author} {\bibinfo {author} {\bibfnamefont {T.~J.}\ \bibnamefont
  {Evans}}, \bibinfo {author} {\bibfnamefont {A.}~\bibnamefont {Schlaus}},
  \bibinfo {author} {\bibfnamefont {Y.}~\bibnamefont {Fu}}, \bibinfo {author}
  {\bibfnamefont {X.}~\bibnamefont {Zhong}}, \bibinfo {author} {\bibfnamefont
  {T.~L.}\ \bibnamefont {Atallah}}, \bibinfo {author} {\bibfnamefont {M.~S.}\
  \bibnamefont {Spencer}}, \bibinfo {author} {\bibfnamefont {L.~E.}\
  \bibnamefont {Brus}}, \bibinfo {author} {\bibfnamefont {S.}~\bibnamefont
  {Jin}}, \ and\ \bibinfo {author} {\bibfnamefont {X.-Y.}\ \bibnamefont
  {Zhu}},\ }\href {\doibase 10.1002/adom.201700982} {\bibfield  {journal}
  {\bibinfo  {journal} {Advanced Optical Materials}\ }\textbf {\bibinfo
  {volume} {6}},\ \bibinfo {pages} {1700982} (\bibinfo {year}
  {2018})}\BibitemShut {NoStop}%
\bibitem [{\citenamefont {Zhang}\ \emph {et~al.}(2017)\citenamefont {Zhang},
  \citenamefont {Shang}, \citenamefont {Du}, \citenamefont {Shi}, \citenamefont
  {Wu}, \citenamefont {Mi}, \citenamefont {Chen}, \citenamefont {Liu},
  \citenamefont {Li}, \citenamefont {Liu}, \citenamefont {Zhang},\ and\
  \citenamefont {Liu}}]{Zhang2017}%
  \BibitemOpen
  \bibfield  {author} {\bibinfo {author} {\bibfnamefont {S.}~\bibnamefont
  {Zhang}}, \bibinfo {author} {\bibfnamefont {Q.}~\bibnamefont {Shang}},
  \bibinfo {author} {\bibfnamefont {W.}~\bibnamefont {Du}}, \bibinfo {author}
  {\bibfnamefont {J.}~\bibnamefont {Shi}}, \bibinfo {author} {\bibfnamefont
  {Z.}~\bibnamefont {Wu}}, \bibinfo {author} {\bibfnamefont {Y.}~\bibnamefont
  {Mi}}, \bibinfo {author} {\bibfnamefont {J.}~\bibnamefont {Chen}}, \bibinfo
  {author} {\bibfnamefont {F.}~\bibnamefont {Liu}}, \bibinfo {author}
  {\bibfnamefont {Y.}~\bibnamefont {Li}}, \bibinfo {author} {\bibfnamefont
  {M.}~\bibnamefont {Liu}}, \bibinfo {author} {\bibfnamefont {Q.}~\bibnamefont
  {Zhang}}, \ and\ \bibinfo {author} {\bibfnamefont {X.}~\bibnamefont {Liu}},\
  }\href {\doibase 10.1002/adom.201701032} {\bibfield  {journal} {\bibinfo
  {journal} {Advanced Optical Materials}\ }\textbf {\bibinfo {volume} {6}},\
  \bibinfo {pages} {1701032} (\bibinfo {year} {2017})}\BibitemShut {NoStop}%
\bibitem [{\citenamefont {Cadelano}\ \emph {et~al.}(2015)\citenamefont
  {Cadelano}, \citenamefont {Sarritzu}, \citenamefont {Sestu}, \citenamefont
  {Marongiu}, \citenamefont {Chen}, \citenamefont {Piras}, \citenamefont
  {Corpino}, \citenamefont {Carbonaro}, \citenamefont {Quochi}, \citenamefont
  {Saba} \emph {et~al.}}]{cadelano2015can}%
  \BibitemOpen
  \bibfield  {author} {\bibinfo {author} {\bibfnamefont {M.}~\bibnamefont
  {Cadelano}}, \bibinfo {author} {\bibfnamefont {V.}~\bibnamefont {Sarritzu}},
  \bibinfo {author} {\bibfnamefont {N.}~\bibnamefont {Sestu}}, \bibinfo
  {author} {\bibfnamefont {D.}~\bibnamefont {Marongiu}}, \bibinfo {author}
  {\bibfnamefont {F.}~\bibnamefont {Chen}}, \bibinfo {author} {\bibfnamefont
  {R.}~\bibnamefont {Piras}}, \bibinfo {author} {\bibfnamefont
  {R.}~\bibnamefont {Corpino}}, \bibinfo {author} {\bibfnamefont {C.~M.}\
  \bibnamefont {Carbonaro}}, \bibinfo {author} {\bibfnamefont {F.}~\bibnamefont
  {Quochi}}, \bibinfo {author} {\bibfnamefont {M.}~\bibnamefont {Saba}},  \emph
  {et~al.},\ }\href {\doibase 10.1002/adom.201500229} {\bibfield  {journal}
  {\bibinfo  {journal} {Advanced Optical Materials}\ }\textbf {\bibinfo
  {volume} {3}},\ \bibinfo {pages} {1557} (\bibinfo {year} {2015})}\BibitemShut
  {NoStop}%
\bibitem [{\citenamefont {Lidzey}\ \emph {et~al.}(2002)\citenamefont {Lidzey},
  \citenamefont {Fox}, \citenamefont {Rahn}, \citenamefont {Skolnick},
  \citenamefont {Agranovich},\ and\ \citenamefont {Walker}}]{Lidzey2002}%
  \BibitemOpen
  \bibfield  {author} {\bibinfo {author} {\bibfnamefont {D.~G.}\ \bibnamefont
  {Lidzey}}, \bibinfo {author} {\bibfnamefont {A.~M.}\ \bibnamefont {Fox}},
  \bibinfo {author} {\bibfnamefont {M.~D.}\ \bibnamefont {Rahn}}, \bibinfo
  {author} {\bibfnamefont {M.~S.}\ \bibnamefont {Skolnick}}, \bibinfo {author}
  {\bibfnamefont {V.~M.}\ \bibnamefont {Agranovich}}, \ and\ \bibinfo {author}
  {\bibfnamefont {S.}~\bibnamefont {Walker}},\ }\href {\doibase
  10.1103/physrevb.65.195312} {\bibfield  {journal} {\bibinfo  {journal}
  {Physical Review B}\ }\textbf {\bibinfo {volume} {65}} (\bibinfo {year}
  {2002}),\ 10.1103/physrevb.65.195312}\BibitemShut {NoStop}%
\bibitem [{\citenamefont {Flatten}\ \emph {et~al.}(2016)\citenamefont
  {Flatten}, \citenamefont {Christodoulou}, \citenamefont {Patel},
  \citenamefont {Buccheri}, \citenamefont {Coles}, \citenamefont {Reid},
  \citenamefont {Taylor}, \citenamefont {Moreels},\ and\ \citenamefont
  {Smith}}]{Flatten2016}%
  \BibitemOpen
  \bibfield  {author} {\bibinfo {author} {\bibfnamefont {L.~C.}\ \bibnamefont
  {Flatten}}, \bibinfo {author} {\bibfnamefont {S.}~\bibnamefont
  {Christodoulou}}, \bibinfo {author} {\bibfnamefont {R.~K.}\ \bibnamefont
  {Patel}}, \bibinfo {author} {\bibfnamefont {A.}~\bibnamefont {Buccheri}},
  \bibinfo {author} {\bibfnamefont {D.~M.}\ \bibnamefont {Coles}}, \bibinfo
  {author} {\bibfnamefont {B.~P.~L.}\ \bibnamefont {Reid}}, \bibinfo {author}
  {\bibfnamefont {R.~A.}\ \bibnamefont {Taylor}}, \bibinfo {author}
  {\bibfnamefont {I.}~\bibnamefont {Moreels}}, \ and\ \bibinfo {author}
  {\bibfnamefont {J.~M.}\ \bibnamefont {Smith}},\ }\href {\doibase
  10.1021/acs.nanolett.6b03433} {\bibfield  {journal} {\bibinfo  {journal}
  {Nano Letters}\ }\textbf {\bibinfo {volume} {16}},\ \bibinfo {pages} {7137}
  (\bibinfo {year} {2016})}\BibitemShut {NoStop}%
\bibitem [{\citenamefont {Virgili}\ \emph {et~al.}(2011)\citenamefont
  {Virgili}, \citenamefont {Coles}, \citenamefont {Adawi}, \citenamefont
  {Clark}, \citenamefont {Michetti}, \citenamefont {Rajendran}, \citenamefont
  {Brida}, \citenamefont {Polli}, \citenamefont {Cerullo},\ and\ \citenamefont
  {Lidzey}}]{Virgili2011}%
  \BibitemOpen
  \bibfield  {author} {\bibinfo {author} {\bibfnamefont {T.}~\bibnamefont
  {Virgili}}, \bibinfo {author} {\bibfnamefont {D.}~\bibnamefont {Coles}},
  \bibinfo {author} {\bibfnamefont {A.~M.}\ \bibnamefont {Adawi}}, \bibinfo
  {author} {\bibfnamefont {C.}~\bibnamefont {Clark}}, \bibinfo {author}
  {\bibfnamefont {P.}~\bibnamefont {Michetti}}, \bibinfo {author}
  {\bibfnamefont {S.~K.}\ \bibnamefont {Rajendran}}, \bibinfo {author}
  {\bibfnamefont {D.}~\bibnamefont {Brida}}, \bibinfo {author} {\bibfnamefont
  {D.}~\bibnamefont {Polli}}, \bibinfo {author} {\bibfnamefont
  {G.}~\bibnamefont {Cerullo}}, \ and\ \bibinfo {author} {\bibfnamefont
  {D.~G.}\ \bibnamefont {Lidzey}},\ }\href {\doibase
  10.1103/physrevb.83.245309} {\bibfield  {journal} {\bibinfo  {journal}
  {Physical Review B}\ }\textbf {\bibinfo {volume} {83}} (\bibinfo {year}
  {2011}),\ 10.1103/physrevb.83.245309}\BibitemShut {NoStop}%
\bibitem [{\citenamefont {Christmann}\ \emph {et~al.}(2008)\citenamefont
  {Christmann}, \citenamefont {Butt{\'{e}}}, \citenamefont {Feltin},
  \citenamefont {Mouti}, \citenamefont {Stadelmann}, \citenamefont {Castiglia},
  \citenamefont {Carlin},\ and\ \citenamefont {Grandjean}}]{Christmann2008a}%
  \BibitemOpen
  \bibfield  {author} {\bibinfo {author} {\bibfnamefont {G.}~\bibnamefont
  {Christmann}}, \bibinfo {author} {\bibfnamefont {R.}~\bibnamefont
  {Butt{\'{e}}}}, \bibinfo {author} {\bibfnamefont {E.}~\bibnamefont {Feltin}},
  \bibinfo {author} {\bibfnamefont {A.}~\bibnamefont {Mouti}}, \bibinfo
  {author} {\bibfnamefont {P.~A.}\ \bibnamefont {Stadelmann}}, \bibinfo
  {author} {\bibfnamefont {A.}~\bibnamefont {Castiglia}}, \bibinfo {author}
  {\bibfnamefont {J.-F.}\ \bibnamefont {Carlin}}, \ and\ \bibinfo {author}
  {\bibfnamefont {N.}~\bibnamefont {Grandjean}},\ }\href {\doibase
  10.1103/physrevb.77.085310} {\bibfield  {journal} {\bibinfo  {journal}
  {Physical Review B}\ }\textbf {\bibinfo {volume} {77}} (\bibinfo {year}
  {2008}),\ 10.1103/physrevb.77.085310}\BibitemShut {NoStop}%
\bibitem [{\citenamefont {Lai}\ \emph {et~al.}(2013)\citenamefont {Lai},
  \citenamefont {Lan},\ and\ \citenamefont {Lu}}]{Lai2013}%
  \BibitemOpen
  \bibfield  {author} {\bibinfo {author} {\bibfnamefont {Y.-Y.}\ \bibnamefont
  {Lai}}, \bibinfo {author} {\bibfnamefont {Y.-P.}\ \bibnamefont {Lan}}, \ and\
  \bibinfo {author} {\bibfnamefont {T.-C.}\ \bibnamefont {Lu}},\ }\href
  {\doibase 10.1038/lsa.2013.32} {\bibfield  {journal} {\bibinfo  {journal}
  {Light: Science {\&} Applications}\ }\textbf {\bibinfo {volume} {2}},\
  \bibinfo {pages} {e76} (\bibinfo {year} {2013})}\BibitemShut {NoStop}%
\bibitem [{\citenamefont {Tassone}\ \emph {et~al.}(1997)\citenamefont
  {Tassone}, \citenamefont {Piermarocchi}, \citenamefont {Savona},
  \citenamefont {Quattropani},\ and\ \citenamefont
  {Schwendimann}}]{Tassone1997}%
  \BibitemOpen
  \bibfield  {author} {\bibinfo {author} {\bibfnamefont {F.}~\bibnamefont
  {Tassone}}, \bibinfo {author} {\bibfnamefont {C.}~\bibnamefont
  {Piermarocchi}}, \bibinfo {author} {\bibfnamefont {V.}~\bibnamefont
  {Savona}}, \bibinfo {author} {\bibfnamefont {A.}~\bibnamefont {Quattropani}},
  \ and\ \bibinfo {author} {\bibfnamefont {P.}~\bibnamefont {Schwendimann}},\
  }\href {\doibase 10.1103/physrevb.56.7554} {\bibfield  {journal} {\bibinfo
  {journal} {Physical Review B}\ }\textbf {\bibinfo {volume} {56}},\ \bibinfo
  {pages} {7554} (\bibinfo {year} {1997})}\BibitemShut {NoStop}%
\bibitem [{\citenamefont {Zuniga-Perez}\ \emph {et~al.}(2014)\citenamefont
  {Zuniga-Perez}, \citenamefont {Mallet}, \citenamefont {Hahe}, \citenamefont
  {Rashid}, \citenamefont {Bouchoule}, \citenamefont {Brimont}, \citenamefont
  {Disseix}, \citenamefont {Duboz}, \citenamefont {Gomm{\'{e}}}, \citenamefont
  {Guillet}, \citenamefont {Jamadi}, \citenamefont {Lafosse}, \citenamefont
  {Leroux}, \citenamefont {Leymarie}, \citenamefont {Li}, \citenamefont
  {R{\'{e}}veret},\ and\ \citenamefont {Semond}}]{Zuniga-Perez2014}%
  \BibitemOpen
  \bibfield  {author} {\bibinfo {author} {\bibfnamefont {J.}~\bibnamefont
  {Zuniga-Perez}}, \bibinfo {author} {\bibfnamefont {E.}~\bibnamefont
  {Mallet}}, \bibinfo {author} {\bibfnamefont {R.}~\bibnamefont {Hahe}},
  \bibinfo {author} {\bibfnamefont {M.~J.}\ \bibnamefont {Rashid}}, \bibinfo
  {author} {\bibfnamefont {S.}~\bibnamefont {Bouchoule}}, \bibinfo {author}
  {\bibfnamefont {C.}~\bibnamefont {Brimont}}, \bibinfo {author} {\bibfnamefont
  {P.}~\bibnamefont {Disseix}}, \bibinfo {author} {\bibfnamefont {J.~Y.}\
  \bibnamefont {Duboz}}, \bibinfo {author} {\bibfnamefont {G.}~\bibnamefont
  {Gomm{\'{e}}}}, \bibinfo {author} {\bibfnamefont {T.}~\bibnamefont
  {Guillet}}, \bibinfo {author} {\bibfnamefont {O.}~\bibnamefont {Jamadi}},
  \bibinfo {author} {\bibfnamefont {X.}~\bibnamefont {Lafosse}}, \bibinfo
  {author} {\bibfnamefont {M.}~\bibnamefont {Leroux}}, \bibinfo {author}
  {\bibfnamefont {J.}~\bibnamefont {Leymarie}}, \bibinfo {author}
  {\bibfnamefont {F.}~\bibnamefont {Li}}, \bibinfo {author} {\bibfnamefont
  {F.}~\bibnamefont {R{\'{e}}veret}}, \ and\ \bibinfo {author} {\bibfnamefont
  {F.}~\bibnamefont {Semond}},\ }\href {\doibase 10.1063/1.4884120} {\bibfield
  {journal} {\bibinfo  {journal} {Applied Physics Letters}\ }\textbf {\bibinfo
  {volume} {104}},\ \bibinfo {pages} {241113} (\bibinfo {year}
  {2014})}\BibitemShut {NoStop}%
\end{thebibliography}%


%merlin.mbs apsrev4-1.bst 2010-07-25 4.21a (PWD, AO, DPC) hacked
%Control: key (0)
%Control: author (8) initials jnrlst
%Control: editor formatted (1) identically to author
%Control: production of article title (-1) disabled
%Control: page (0) single
%Control: year (1) truncated
%Control: production of eprint (0) enabled
%

\end{document}